\documentclass[journal,draftcls,onecolumn,12pt]{IEEEtran}
\usepackage{graphicx}
\usepackage{amsmath}
\usepackage{caption}
\usepackage{subcaption}
\usepackage{epsfig}
\usepackage{float}
\usepackage{lipsum}
\usepackage{stfloats}
\usepackage{array}
\usepackage{amssymb}
\usepackage{mathtools}
\usepackage{cite}
\usepackage{url}
\usepackage{bm}
\usepackage{amsthm}
\usepackage{algorithm2e}
\usepackage{algorithmic}
\usepackage{multirow}
\usepackage{fancyh}
\usepackage[utf8]{inputenc}
\usepackage{bbm}
\usepackage{dsfont}
\usepackage{upgreek}
\usepackage{stmaryrd}
\normalsize
\IEEEoverridecommandlockouts

\theoremstyle{plain}

\newtheorem{proposition}{Proposition}

\theoremstyle{definition}
\newtheorem{example}{Example}

\makeatletter
\DeclareRobustCommand{\cev}[1]{%
	\mathpalette\do@cev{#1}%
}
\newcommand{\do@cev}[2]{%
	\fix@cev{#1}{+}%
	\reflectbox{$\m@th#1\vec{\reflectbox{$\fix@cev{#1}{-}\m@th#1#2\fix@cev{#1}{+}$}}$}%
	\fix@cev{#1}{-}%
}
\newcommand{\fix@cev}[2]{%
	\ifx#1\displaystyle
	\mkern#23mu
	\else
	\ifx#1\textstyle
	\mkern#23mu
	\else
	\ifx#1\scriptstyle
	\mkern#22mu
	\else
	\mkern#22mu
	\fi
	\fi
	\fi
}

\newcommand{\vect}[1]{\bm{#1}}

\DeclareMathOperator*{\1}{\mathds{1}}

\newcommand{\bv}{{\vect{v}}}

\newcommand{\Phib}{{\Phi}_{\rm  a}}
\newcommand{\Phibact}{{\tilde{\Phi}}_{\rm  a}}
\newcommand{\Phiu}{{\Phi}_{\rm  u}}

\newcommand{\Phiuq}{{\Phi}_{\rm u}^{(q)}}

\newcommand{\Phiuqact}{{\tilde{\Phi}}_{\rm u}^{(q)}}

\newcommand{\lambdab}{{\lambda}_{\rm  a}}
\newcommand{\lambdabact}{{\tilde{\lambda}}_{\rm  a}}
\newcommand{\lambdau}{{\lambda}_{\rm  u}}
\newcommand{\lambdauq}{{\lambda}}
\newcommand{\lambdauqact}{{\tilde{\lambda}}}

\newcommand{\Rbb}{{\mathbb{R}}}

\newcommand{\Bc}{B_{\rm \scriptscriptstyle c}}

\newcommand{\Np}{N_{\rm p}}

\newcommand{\Rin}{R_{\rm  in}}

\newcommand{\Rout}{R_{\rm out}}
\newcommand{\Acal}{{\mathcal{A}}}

\newcommand{\Bcal}{{\mathcal{B}}}

\newcommand{\Vcal}{{\mathcal{V}}}

\newcommand{\Ebb}{{\mathbb{E}}}
\newcommand{\Pbb}{{\mathbb{P}}}
\newcommand{\Dd}{{\mathrm{d}}}
\newcommand{\pnot}{{p_0}}
\newcommand{\pone}{{p_1}}
\newcommand{\pu}{{p_{\rm u}}}
\newcommand{\pact}{{p_{\rm a}}}

\newcommand{\sigmaNtwo}{\sigma^2_{\rm \scriptscriptstyle N}}
\newcommand{\SIR}{{\mathrm{SIR}}}
\newcommand{\sigmaStwo}{\sigma^2_{\rm \scriptscriptstyle S}}
\newcommand{\sigmaItwo}{\sigma^2_{\rm \scriptscriptstyle I}}
\newcommand{\Pu}{P_{\rm u}}

\newcommand{\ak}{a_k}
\newcommand{\ao}{a_{0}}
\newcommand{\aone}{a_{1}}
\newcommand{\uo}{u_{0}}
\newcommand{\uj}{u_{j}}

\newcommand{\Cpilot}{{\bar{C}}^{\rm \scriptscriptstyle pilot}}

\newcommand{\Cbaro}{{\bar{R}}_0}

\newcommand{\deltao}{\delta_0}

\newcommand{\Careacm}{{\mathcal{C}}^{\rm \scriptscriptstyle  cellular}}
\newcommand{\Careafog}{{\mathcal{C}}^{\rm \scriptscriptstyle  fog}}
\newcommand{\roone}{r_{1,0}}
\newcommand{\roo}{r_{0,0}}

\newcommand{\ThreSig}{\tau_{\rm useful}}
\newcommand{\ThreInt}{\tau_{\rm interf}}
\newcommand{\Pstrfog}{P_{\rm s}^{\rm fog}}
\newcommand{\Pstrcell}{P_{\rm s}^{\rm cell}}
\newcommand{\Pavg}{P_{\rm a}}

\newcommand{\Cjinfty}{C_{j}^{\infty}}
\newcommand{\Coinfty}{C_{0}^{\infty}}
\newcommand{\Cbaroinfty}{{\bar{C}}_{0}^{\infty}}

\allowdisplaybreaks
	
\usepackage{setspace}	

\newfont{\bbb}{msbm10 scaled 700}

\newfont{\bb}{msbm10 scaled 1100}
\newcommand{\CC}{\mbox{\bb C}}
\newcommand{\PP}{\mbox{\bb P}}
\newcommand{\RR}{\mbox{\bb R}}

\newcommand{\EE}{\mbox{\bb E}}

\newcommand{\ev}{{\bf e}}

\newcommand{\gv}{{\bf g}}
\newcommand{\hv}{{\bf h}}

\newcommand{\sv}{{\bf s}}

\newcommand{\wv}{{\bf w}}
\newcommand{\xv}{{\bf x}}
\newcommand{\yv}{{\bf y}}
\newcommand{\zv}{{\bf z}}
\newcommand{\zerov}{{\bf 0}}
\newcommand{\onev}{{\bf 1}}

\newcommand{\Gm}{{\bf G}}

\newcommand{\Ym}{{\bf Y}}
\newcommand{\Zm}{{\bf Z}}

\newcommand{\Ac}{{\cal A}}
\newcommand{\Cc}{{\cal C}}

\newcommand{\Gc}{{\cal G}}

\newcommand{\Nc}{{\cal N}}

\newcommand{\Sc}{{\cal S}}
\newcommand{\Uc}{{\cal U}}
\newcommand{\Wc}{{\cal W}}
\newcommand{\Vc}{{\cal V}}
\newcommand{\Xc}{{\cal X}}

\renewcommand{\Re}{{\rm Re}}

\newcommand{\herm}{{\sf H}}

\newcommand{\transp}{{\sf T}}
\renewcommand{\vec}{{\rm vec}}

\usepackage{hyperref}
\hypersetup{
    bookmarks=true,         
    unicode=false,          
    pdftoolbar=true,        
    pdfmenubar=true,        
    pdffitwindow=false,     
    pdfstartview={FitH},    
    pdfnewwindow=true,      
    colorlinks=true,       
    linkcolor=red,          
    citecolor=green,        
    filecolor=blue,      
    urlcolor=blue           
}

\begin{document}
\title{Fog Massive MIMO: A User-Centric Seamless Hot-Spot Architecture}
\author{Ozgun Y. Bursalioglu$^\dagger$, Giuseppe Caire$^\star$, Ratheesh K. Mungara$^\star$, \\
Haralabos C. Papadopoulos$^\dagger$, Chenwei Wang$^\dagger$
\thanks{$^\dagger$ Docomo Innovations, Inc., Palo Alto, CA 94304, USA (Email: \{obursalioglu, hpapadopoulos, cwang\}@docomoinnovations.com).
$^\star$ Communications and Information Theory Group, Technische Universit{\"a}t Berlin, Berlin 14059, Germany (Email: \{caire, mungara\}@tu-berlin.de). 
}
}

\maketitle

\begin{abstract}
\linespread{1.3}\small
The decoupling of data and control planes, as proposed for 5G networks, will enable
the efficient implementation of multitier networks where user equipment (UE) nodes 
obtain coverage and connectivity through the {\em top-tier} macro-cells,
and, at the same time,  achieve high-throughput low-latency communication  through lower tiers in the hierarchy. 
This paper considers a new architecture for such lower tiers, dubbed {\em fog massive MIMO}, where 
the UEs are able to establish high-throughput low-latency data links in a seamless and opportunistic manner, 
as they travel through a dense ``fog'' of  high-capacity wireless infrastructure nodes, referred to as remote radio heads (RRHs). 
Traditional handover mechanisms in dense multicell networks inherently give rise to frequent handovers and pilot sequence re-assignments, incurring, as a result, excessive protocol overhead and significant latency.
In the proposed fog massive MIMO architecture, UEs seamlessly and implicitly associate themselves 
to the most convenient RRHs in a completely autonomous manner. 
Each UE makes use of a unique uplink pilot sequence, and pilot contamination is mitigated by a 
novel  coded ``on-the-fly'' pilot contamination control mechanism. 
We analyze the spectral efficiency and the outage probability of the proposed architecture via stochastic geometry, using some recent results on {\em unique coverage in Boolean models}, and provide a  detailed comparison with respect to an idealized baseline massive MIMO cellular system, that neglects protocol overhead and latency due to explicit user-cell association. 
Our analysis, supported by extensive system simulation, reveals that there exists a ``sweet spot'' of  the per-pilot user load (number of users per pilot),  
such that the proposed system achieves  spectral efficiency close to that of an ideal 
cellular system with the minimum distance user-base station association 
and no pilot/handover overhead.
\end{abstract}


\IEEEpeerreviewmaketitle


\section{Introduction}
\label{sec:intro}


5G technologies are expected to bring about great improvements with respect to a multitude of metrics, including user and cell throughput, end-to-end latency and massive device connectivity.
They are also viewed as essential elements for enabling the much broader gamut of services envisioned, such as immersive applications (e.g., virtual/augmented/mixed reality) \cite{bastug2017toward,qi2016quantifying},  
haptics \cite{aijaz2017realizing}, V2X (vehicle-to-everything) \cite{luoto2016system}, and the Internet of Things \cite{mavromoustakis2016internet}. To meet such ambitious and broad range of goals, operators would have to rely on a combination of additional resources, which include
newly available licensed and unlicensed bands, network densification, large antenna arrays, and new PHY/network layer technologies.
In particular, the wide range of performance objectives for such a disparate variety of services is not suited to the conventional
``one-size fits all'' approach of conventional single-tier cellular networks. 
For this reason, a key feature of 5G networks consists of the decoupling of data and control planes \cite{Boccardi-5G}, in order to 
enable multi-tier networks \cite{hossain2014evolution}.  In such architectures, user equipment (UE) nodes shall maintain 
coverage and connectivity through macro-cells operating  at conventional frequency bands (1 to 3.5 GHz), with low propagation pathloss and good indoor penetration. While this tier stays at the top of the hierarchy and takes care of fundamental functionalities such as mobility management and general bookkeeping of the users in the system, high-throughput and low-latency data communication is supported by lower tiers in the hierarchy, formed by smaller and simpler infrastructure nodes operating at higher frequency bands with smaller range \cite{ishii2012novel}.  
By means of a large number of remote radio heads (RRHs), deployed as a second tier, 
localized network densification can be used for hot-spot formation, in order to tackle the spatial non-uniformity in data traffic demands, 
which is one of the biggest challenges faced by wireless operators.\footnote{A recent study revealed that 90\% of the data 
is consumed by 10\% of the users within 5\% of the area~\cite{MWC-Viavi}.}


Massive MIMO suppresses the small-scale fading effects through {\em channel hardening} \cite{Marzetta-massiveMIMO}, 
resulting in an almost-deterministic wireless channel between the transmitter and receiver, which in turn 
simplifies rate and power allocation and yields superior spectral and energy efficiency. 
While massive MIMO has been mainly regarded as a technology for large and costly cellular base stations (BSs), 
the current technology trend considers higher and higher carrier frequencies and mass production for dense deployment, 
with corresponding decreasing size and costs. It is therefore expected that, in the near future, it will be possible to implement 
small and inexpensive RRHs with up to $\sim$100 antennas, each of which serving on average 
a relatively small number of users (e.g., $\sim$10) on same channel subband. 
Considering the fact that the main deployment cost for operators is represented by the number of sites (not number of antennas per site), 
to make the most out of BS sites, we focus on RRH with large antenna arrays, unlike single antenna site work such as~\cite{CellfreevsSmallcells}.

Higher-frequency operation inherently implies shorter range,  
higher penetration loss, and shorter coherence time.  
Higher-frequency tiers would need to be denser.
As a result, planned operation of such tiers becomes excessively costly and inefficient.  
Indeed, in a conventional small-cell architecture, the cell size is simply scaled down, 
while maintaining all functionalities in each small-cell BS. The performance of dense small-cell systems is therefore severely limited 
by several issues resulting from cell-size scale down, e.g., reduced cell isolation, excessive handoffs, 
asymmetric forward and reverse links, dynamic interference, and backhaul management~\cite{HenNet7ways}. 
Particularly, in the presence of high mobility (as for V2X applications),
frequent handovers and per-cell pilot sequence re-assignment may incur a large protocol overhead and significant latency. 
For example, when a finite number of mutually orthogonal pilot sequences are distributed among 
the users and are not re-allocated dynamically,  it is unavoidable that the same pilot may be assigned to several users that may be in spatial 
proximity at some point in time (due to mobility). 
In this case, a dense massive MIMO deployment may incur severe multiuser interference due to 
{\em pilot contamination} \cite{pilot-contam-MC}.
%
%

\subsection{Contributions}  \label{contributions}

This paper considers a new architecture for dense massive MIMO, dubbed {\em fog massive MIMO}, where 
the UEs are able to establish high-throughput and low-latency data links in a seamless and opportunistic manner, 
as they move through a dense ``fog'' of high-capacity multiantenna RRHs.
Our architecture follows a user-centric approach (see also our preliminary work in \cite{on-the-fly-ICC16, coded-pilots}). In particular, fog massive MIMO
realizes the advantages of distributed antenna systems such as cell-free operations and efficient interference 
management, while preserving the simplicity of {\em localized} (i.e., per-RRH) physical layer processing.
This is achieved by exploiting the channel state information obtained via time-division duplexing (TDD) 
uplink (UL) downlink (DL) reciprocity, in conjunction with two mechanisms: \emph{(i)} {\em On-the-fly pilot contamination control} at the physical layer; 
\emph{(ii)} {\em Geographic routing with packet replication} at the network layer. 

The proposed on-the-fly pilot contamination control mechanism allows
each RRH to decide autonomously and instantaneously whether a received UL pilot is free from contamination. 
Only the packets of the UEs corresponding to uncontaminated pilots  are decoded in the UL slot and subsequently served via multiuser MIMO precoding in the DL slot.

Geographic routing with packet replication allows any user, with high probability, to be in the proximity of 
a sufficiently large number of RRHs with data packets ready to send, thus creating opportunistic macro-diversity opportunities.  
Hence, it is not really important ``from which RRH'' a given UE receives a requested packet, as long as there exists some RRHs nearby 
that can provide such packet. 

\begin{figure}
	\centering
	{
		\begin{minipage}[b]{.45\linewidth}
			\centering
			{
				\includegraphics[scale=0.24]{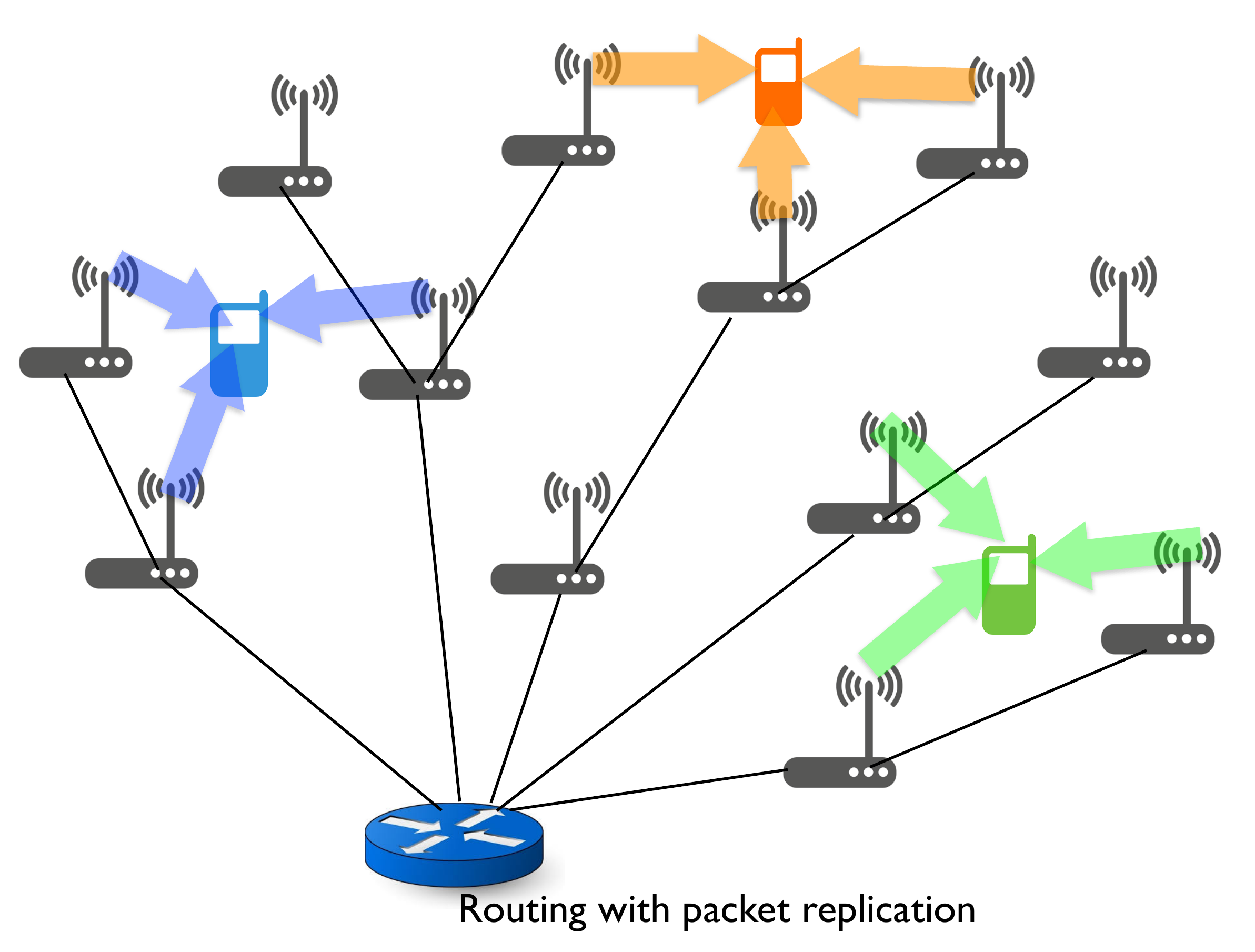} 
			}
			\subcaption{ \label{subfig:fogconcept1}}
		\end{minipage}
		\begin{minipage}[b]{.45\linewidth}
			\centering
			{
				\includegraphics[scale=0.24]{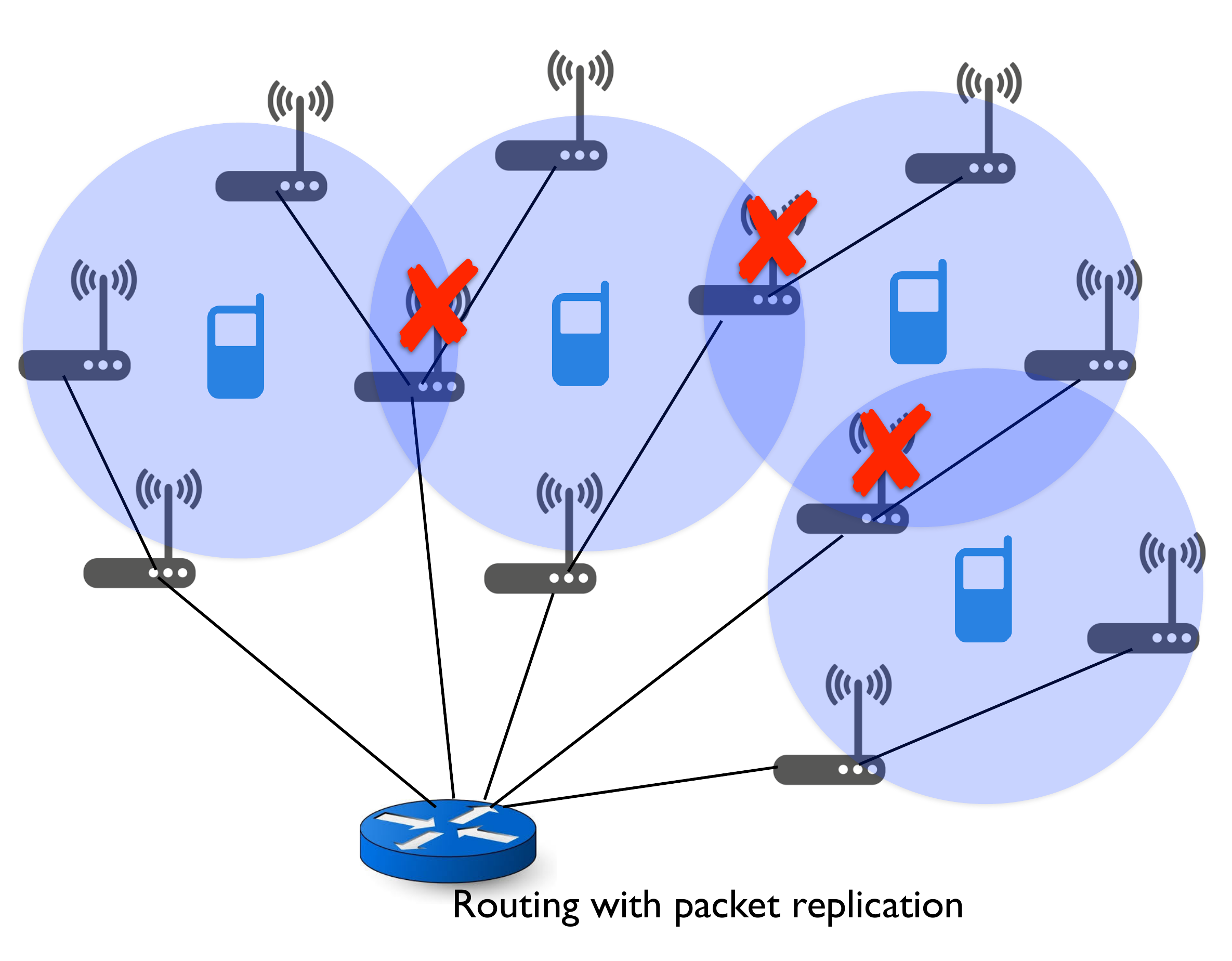} 
			}
			\subcaption{ \label{subfig:fogconcept2}}
		\end{minipage}
		\caption{\small
A fog massive MIMO network serving three UEs. In example (a), the UEs use orthogonal UL pilots, and receive their 
requested packet from possibly multiple RRHs (macro-diversity). Example (b) shows three UEs associated to the same UL pilot.
The RRHs in the transmission range intersection regions detect pilot contamination and discard the corresponding channel estimate.}
		\label{Fig:fogconcept}
	}
\end{figure}

Fig.~\ref{Fig:fogconcept} shows the concept of the proposed fog massive MIMO system. In Fig.~\ref{subfig:fogconcept1}, three
UEs making use of orthogonal UL pilots  receive their requested DL packet from some surrounding RRHs, which beamform 
the packet such that the signal combine coherently, achieving a macro-diversity effect. 
In Fig.~\ref{subfig:fogconcept2} we illustrate the idea of on-the-fly pilot contamination control.
In this case, the three UEs use the same UL pilot. Hence, some RRHs in the intersection of their transmission range suffer from contamination.
These RRHs refrain from using this pilot for channel estimation and beamforming. 
Nevertheless, the UEs can still be served by other RRHs.


This altogether motivates the analysis of a fog massive MIMO system that we present here, with the following novel contributions:

1) We propose a simple {\em coded UL pilot scheme} that detects with arbitrarily high reliability (in the limit of large antennas) the pilot contamination
through a very simple pilot matched filtering and threshold operation. This allows on-the-fly pilot contamination control on a per-slot basis, i.e., with minimal latency.

2) We leverage some recent results on \emph{unique coverage in Boolean models}~\cite{unique-coverage-Haenggi} to provide the performance analysis 
of the proposed fog massive MIMO system when the locations of UEs and RRHs are modeled via two independent Poisson point processes (PPPs).
Our analysis yields the spatially averaged user spectral efficiency (b/s/Hz per user) and
area spectral efficiency (b/s/Hz per unit area) through easy to compute integral approximation, extensively validated by Monte Carlo system simulation.

3) As a term of comparison, we consider an ideal cellular massive MIMO system with minimum distance UE-BS 
association and fractional pilot reuse (each cell uses a randomly selected subset of pilots from a pool of orthogonal UL pilots). 
For this system, we derive the expressions for the spatially averaged user and area spectral efficiencies in integral forms. 

As  expected, the ideal cellular system yields generally better performance than the (overhead-less) fog system, 
since it associates UEs to BSs according to the minimum distance and assumes an ideal intra-cell pilot management policy that guarantees 
mutually orthogonal pilots in each cell. In our comparison, we do not take into account the cost incurred
by pilot allocation/reallocation at each handover in the presence of mobility, when users migrate from cell to cell.
Instead of trying to quantify such overhead, which would be beyond the scope of this paper, we compare the ideal baseline cellular system with the 
fully practical proposed fog massive MIMO system, and identify the regime where the latter yields comparable performance with respect to its 
cellular counterpart.

\subsection{Relation to Cell-Free Massive MIMO}

A system consisting of RRHs with local processing,  geographic routing with packet replication, and 
somehow user-centric operations  was recently proposed in \cite{CellfreevsSmallcells} under the name of {\em cell-free massive MIMO}.
This system makes use of a very large number of single-antenna RRHs. 
The key idea of cell-free massive MIMO is that in the DL, maximal-ratio transmission can be obtained 
by local combining of the user data packets with the complex conjugate of the channel coefficients at each RRH, 
without the need for centralized processing as in 
cloud radio access network (CRAN) architectures (e.g., see \cite{CloudRAN-simeone-Yu}). 
As in our scheme, the channel coefficients can be learned at each RRH via TDD reciprocity from UL pilots. 
Despite some similarities, cell-free massive MIMO and the fog massive MIMO proposed and analyzed in this paper are indeed 
quite different.  First, in cell-free massive MIMO it is assumed that each RRH has full information on the channel 
large-scale gains from itself to all users in the system. Otherwise, for $K$ UEs with $Q$ orthogonal pilots, with $K > Q$, 
it is impossible for a RRH to associate a packet to the channel estimate produced from a given pilot observation, since the same pilot
is associated to multiple users. As an alternative, cell-free massive MIMO considers $K$ non-orthogonal pilots over $Q < K$ dimensions, 
with unique user-pilot association. However, in this case the scheme requires careful power allocation with possibly iterative re-computation of
the non-orthogonal pilot codebook, where again the large-scale gain information for all user-RRH pairs is needed. 
This is quite impractical to be estimated and maintained, especially in a dynamic mobility environment. 
Then, cell-free massive MIMO yields competitive spectral efficiency performance for very high RRH density, 
typically much higher than the UE density, which is again quite impractical. 
Finally, and perhaps more importantly, cell-free massive MIMO  is highly asymmetric between UL and DL. 
In fact, while in the DL maximal ratio transmission can be achieved by local combining at each RRH, in the UL each 
single-antenna RRH must send its received signal to a central processor to enable multiuser multiantenna joint processing,  as done in 
 be concentrated to a single central processor as in CRAN.
Since UL and DL in CRAN have similar complexity and performance (as rigorously stated by duality theorems \cite{liu2016uplink}), 
one may wonder what is the significance  of making the DL very simple at the cost of a large performance degradation with respect to CRAN, 
if a CRAN is anyway needed for the UL. 

Due to the above mentioned critical system assumptions and to the requirement of centralized processing in the UL,  
it is very hard to make a fair comparison between our proposed system and the cell-free massive MIMO system of \cite{CellfreevsSmallcells}. 
For this reason, we have chosen to defer such comparison to some future work and here we compare with a baseline idealized dense 
cellular massive MIMO system (as described before), providing a benchmark for any system based on per-cell processing.



\section{Fog Massive MIMO System}

\label{sec:Net model Fog}

In the considered  fog massive MIMO system, a number of multiantenna RRHs and single-antenna UEs are deployed in a given planar region. 
The system operates in TDD mode, where the UL and the DL take place on the same frequency band, but in
two adjacent subslot intervals within a TDD slot.
It is assumed that the propagation channel coefficients remain constant over a TDD slot, and that physical layer UL/DL reciprocity is 
achieved (e.g., using relative calibration techniques such as those in \cite{shepard2012argos,rogalin2014scalable}).
Then, both the UL and the DL channel coefficients can be learned by the RRHs in each TDD slot from the UL pilot signal sent 
by the UEs in the UL.


Due to low-latency requirements, the TDD slot interval is fixed to be quite short.
%
Letting $T$ denote the number of signal dimensions per slot, in an outdoor scenario with the coherence bandwidth of $200-500$ kHz 
and the slot duration of $1$ ms, we have $T \approx $ 200 to 500 signal dimensions. 
In this paper, we assume that pilot transmission for channel estimation, user scheduling, and data transmission, must take place on a slot-by-slot basis. 
This allows the maximum flexibility to the possible user scheduling algorithms (not considered here), since the scheduled users in each slot may freely change on a per-slot basis. 
The analysis of UL data transmission, triggered by the successful DL decoding, 
is similar and straightforward with respect to its DL counterpart. Thus, we consider a simplified slot structure consisting of UL channel estimation and DL data transmission, without UL data transmission.

The locations of the RRHs $\{ \ak \}$ are modeled as a homogeneous PPP $\Phib \subset \Rbb^2$ of density $\lambdab$, 
implying that on average there are $\lambdab$ RRHs per unit area. The locations of the users $\{ \uj \}$  are modeled as an another 
independent homogeneous PPP $\Phiu \subset \Rbb^2$ of density $\lambdau$. 

\subsection{Coded UL Pilots} \label{sec:coded UL pilots}


Let $L < T$ denote the pilot dimension, i.e., the length of the pilot sequences. 
Assume that $L = Q + Q'$, where $Q$ and $Q'$ are integers, and $Q'$ is even. We consider a particular format of coded pilots defined as follows. 
Let $\Sc_Q = \{ \sv_q \in \CC^Q : q \in [Q]\}$ denote\footnote{where the notation $[n]$ indicates the set of indices $\{1,\ldots, n\}$.}  
a set of $Q$ mutually orthogonal sequences of dimension $Q$ and norm 1.
Also, let $\Wc(Q',Q'/2) = \{ \wv_\ell : \ell  \in [{Q' \choose Q'/2}] \}$ denote the set of all equal-weight binary codewords
of length $Q'$ with Hamming weight equal to $Q'/2$. 
Then, the UL pilot codebook is formed by the set of $L$-dimensional sequences
\begin{equation} \label{codebookC}
\Cc = \left \{ \xv_{q,\ell} = \sqrt{P_{\rm u}} \left [ \sqrt{Q} \sv^\transp_q,  \sqrt{2} \wv^\transp_\ell \right ]^\transp \; : \;  
\forall \; (q,\ell) \in  [Q] \times \left[{Q' \choose Q'/2} \right] \right \}. 
\end{equation} 
It is immediate to verify that  $\|\xv_{q,\ell}\|^2 = P_{\rm u} L$, such that the UL transmit energy per pilot dimension is effectively equal to $P_{\rm u}$. 
Notice also that the pilot codebook $\Cc$ in (\ref{codebookC}) is partitioned into $Q$ groups
$\Cc_q : q \in [Q]$, where $\Cc_q = \{\xv_{q,\ell} : \ell \in [{Q' \choose Q'/2}] \}$, with the
property that two pilot sequences in distinct groups $\Cc_q$ and $\Cc_{q'}$ 
are orthogonal on their first $Q$ components.

Each UE is assigned a pilot sequence in $\Cc$ independently with uniform probability, such that  the user PPP 
can be viewed as $\Phiu = \cup_{q=1}^{Q} \Phiuq$, where  $\Phiuq$ is the process of user locations utilizing pilots in 
group $\Cc_q$, and where $\lambda = \lambdau/Q$ is the corresponding density, common to all pilot groups.
For simplicity, in the analysis we assume that, at each TDD slot, no two users assigned to the same group $q$ 
coincide also in the  second pilot field $\wv_\ell$. In practice, the size of the equal-weight code is large enough, 
such that the probability that two users in the same group $q$ have also the same $\wv_\ell$ is very small. For example, 
setting $Q' = 20$ as in our numerical results yields ${Q' \choose Q'/2} = 184756$.

\subsection{UL Channel Estimation and Pilot Contamination Control}  \label{sec:on-the-fly pilot contam control}

Both the data and the pilot symbols are transmitted as time-frequency symbols assuming OFDM modulation. 
Focusing on a given TDD time slot, 
let $\gv_{k,j} = \sqrt{\beta_{k,j}} \hv_{k,j}$ denote the propagation channel vector between the antenna array of the $k$-th 
RRH located at $a_k$ and the $j$-th user located at $u_j$. The coefficient $\beta_{k,j}$ denotes the large-scale distance dependent channel gain, 
the statistics of which will be defined later. 
The vector $\hv_{k,j} \in \CC^M$ contains the small-scale channel fading coefficients, with IID (independent and identically distributed) components $\sim \Cc\Nc(0,1)$, 
mutually independent across users and RRHs. The received signal at the $k$-th RRH over the first $Q$ components of the UL pilot field is given by the $M \times Q$ array 
\begin{equation} \label{Qpilot1}
\Ym_k^{{\rm pilot},1} = \sqrt{P_{\rm u} Q} \sum_{q' \in [Q]} \left ( \sum_{j : u_j \in  \Phiu^{(q')}}  \gv_{k,j} \right ) \sv^\transp_{q'}  + \Zm_k^{{\rm pilot},1}, 
\end{equation}  
where $\Zm_k$ contains AWGN samples $\sim \Cc\Nc(0,\sigmaNtwo)$ with $\sigmaNtwo$ representing the noise variance. 
After correlating with respect to the pilot sequences \cite{Marzetta-massiveMIMO}, the channel estimates corresponding to each $q$-th pilot group are given by
\begin{equation} \label{qpilot1}
\widehat{\gv}_k^{(q)}  =  \frac{1}{\sqrt{\Pu Q}} \Ym_k^{{\rm pilot},1} \sv_q^* =  \left ( \sum_{j : u_j \in  \Phiuq}  \gv_{k,j} \right )   + \zv_{k,q}^{{\rm pilot},1}, \;\;\; q \in [Q], 
\end{equation}  
where $\zv_{k,q}^{{\rm pilot},1} \in \CC^M$ has IID components $\sim \Cc\Nc(0,\sigmaNtwo/(\Pu Q))$. 
The pilot contamination is due to the fact that the channel estimate
$\widehat{\gv}_k^{(q)}$ contains the superposition of the channels all users in the same pilot group $q$. 
Qualitatively speaking, the pilot observation $\widehat{\gv}_k^{(q)}$ can be ``trusted'' only if it contains a single strong channel contribution. 
In contrast, if there are no or more than one strong contribution,  pilot $q$ is considered as ``untrusted'' and the corresponding channel observation is discarded. 
If pilot $q$ is trusted, then the $k$-th RRH uses  $\widehat{\gv}_k^{(q)}$ to calculate the DL precoding vector and send DL precoded data to the associated (single) strong user.  Otherwise, it will simply ignore the channel estimate $\widehat{\gv}_k^{(q)}$ and refrain from transmission of DL data to any user 
associated to group $q$.  

We next consider {\em how each RRH can distinguish between trusted and untrusted pilot observations in an autonomous  
decentralized manner.} 
This is obtained by exploiting the equal-weight code in the second $Q'$ components of the UL pilot field. The corresponding received signal is given by 
\begin{equation} \label{Qpilot2}
\Ym_k^{{\rm pilot},2} = \sqrt{ 2 P_{\rm u}} \sum_{q' \in [Q]} \left ( \sum_{j : u_j \in  \Phiu^{(q')}}  \gv_{k,j} \wv^\transp_{\ell_j} \right ) + \Zm_k^{{\rm pilot},2}, 
\end{equation}  
where, by construction, the indices $(q',\ell_j)$ are all distinct. 
Performing maximal ratio combining with respect to the channel estimate $\widehat{\gv}_k^{(q)}$, we find 
\begin{eqnarray} 
\yv_{k,q}^{{\rm pilot},2} = \frac{1}{M} \left ( \Ym_k^{{\rm pilot},2} \right )^\herm \widehat{\gv}_k^{(q)} & = & \sqrt{2 \Pu} \sum_{j : u_j \in  \Phiuq}  \frac{1}{M} \| \gv_{k,j} \|^2 \wv_{\ell_j}  \label{useful} \\
&  & + \mbox{signal} \times \mbox{signal} 
+ \mbox{signal} \times \mbox{noise} 
+ \mbox{noise} \times \mbox{noise} \label{garbage}
\label{noisexnoise} 
\end{eqnarray}
As in the classical massive MIMO analysis \cite{Marzetta-massiveMIMO}, it is immediate to show that all terms involving inner products of channel vectors with different indices,  inner product of channel vectors times noise vectors, and noise vectors times noise vectors (terms in (\ref{garbage}))
converge to zero with probability 1 as $M \rightarrow \infty$. Hence, using the fact that $\frac{1}{M} \| \gv_{k,j} \|^2 \rightarrow \beta_{k,j}$ we have that for sufficiently large $M$ the second pilot field after maximal ratio combining is given by 
\begin{equation} \label{qpilot2} 
\yv_{k,q}^{{\rm pilot},2} = \sqrt{2\Pu} \sum_{j : u_j \in  \Phiuq}  \beta_{k,j}  \wv_{\ell_j}   + \zv_{k,q}^{{\rm pilot}, 2} 
\end{equation}
where $\zv_{k,q}^{{\rm pilot}, 2}$ has mean zero and variance $O(1/M)$. 
In brief, the second pilot field yields the weighted sum of the (unique) equal-weight binary codewords for users in pilot group $q$, with weighting coefficients given by 
the corresponding channel large-scale coefficients, plus a small residual noise plus interference that vanishes for large $M$.  
At this point, the decision of whether pilot $q$ is trusted 
or not is obtained by the following simple thresholding mechanism: 

{\bf Trusted pilot detection rule.}
Set two thresholds $\tau_{\rm useful} \geq \tau_{\rm interf} \geq 0$, and define the one-bit threshold quantizer 
$\Delta( \cdot , \tau)$ such that for any real vector $\xv$, $\widehat{\xv} = \Delta( \xv, \tau)$ is the binary vector with components
$\widehat{x}_i =  0$ if $x_i < \tau$ and $\widehat{x}_i =  1$ if $x_i \geq \tau$. 
Then, pilot $q$ is trusted at RRH $k$ if the following two conditions hold: \\
1) $\widehat{\wv}_{k,q} = \Delta(\Re\{\yv_{k,q}^{{\rm pilot},2}\}, \tau_{\rm useful})$ is an equal-weight codeword in $\Wc(Q',Q'/2)$; \\
2) Letting $\odot$ denote elementwise product, $\onev$ be the all-1 vector and $\zerov$ be the all-0 vector, it must be
\begin{equation} \label{interf-rejection}
( \onev -   \widehat{\wv}_{k,q}) \odot \Delta(\Re\{\yv_{k,q}^{{\rm pilot},2}\}, \tau_{\rm interf}) = \zerov. 
\end{equation}

The above rule is explained as follows: if the sum in (\ref{qpilot2}) contains a single strong user such that $\sqrt{2\Pu} \beta_{k,j} > \tau_{\rm useful}$,  
while the sum of the all other terms plus noise is sufficiently weak, then condition 1) above is satisfied, i.e., 
$\widehat{\wv}_{k,q} \in \Wc(Q',Q'/2)$. 
However, some users using pilots from the same group $\Cc_q$, termed co-pilot users,  may yield a contribution barely 
below the threshold $\tau_{\rm useful}$.  In order to control how weaker the co-pilot interference should be with 
respect to the useful signal,  we introduce a second threshold $\tau_{\rm interf}$, which represents an acceptable ``rise over thermal'' level. 
If condition (\ref{interf-rejection}) holds, then in the positions of the ``zeros'' in $\widehat{\wv}_{k,q}$, 
the magnitude of $\Re\{\yv_{k,q}^{{\rm pilot},2}\}$ is below the more restrictive threshold $\tau_{\rm interf}$. 

\begin{figure}[ht]
\centerline{\includegraphics[scale=0.45]{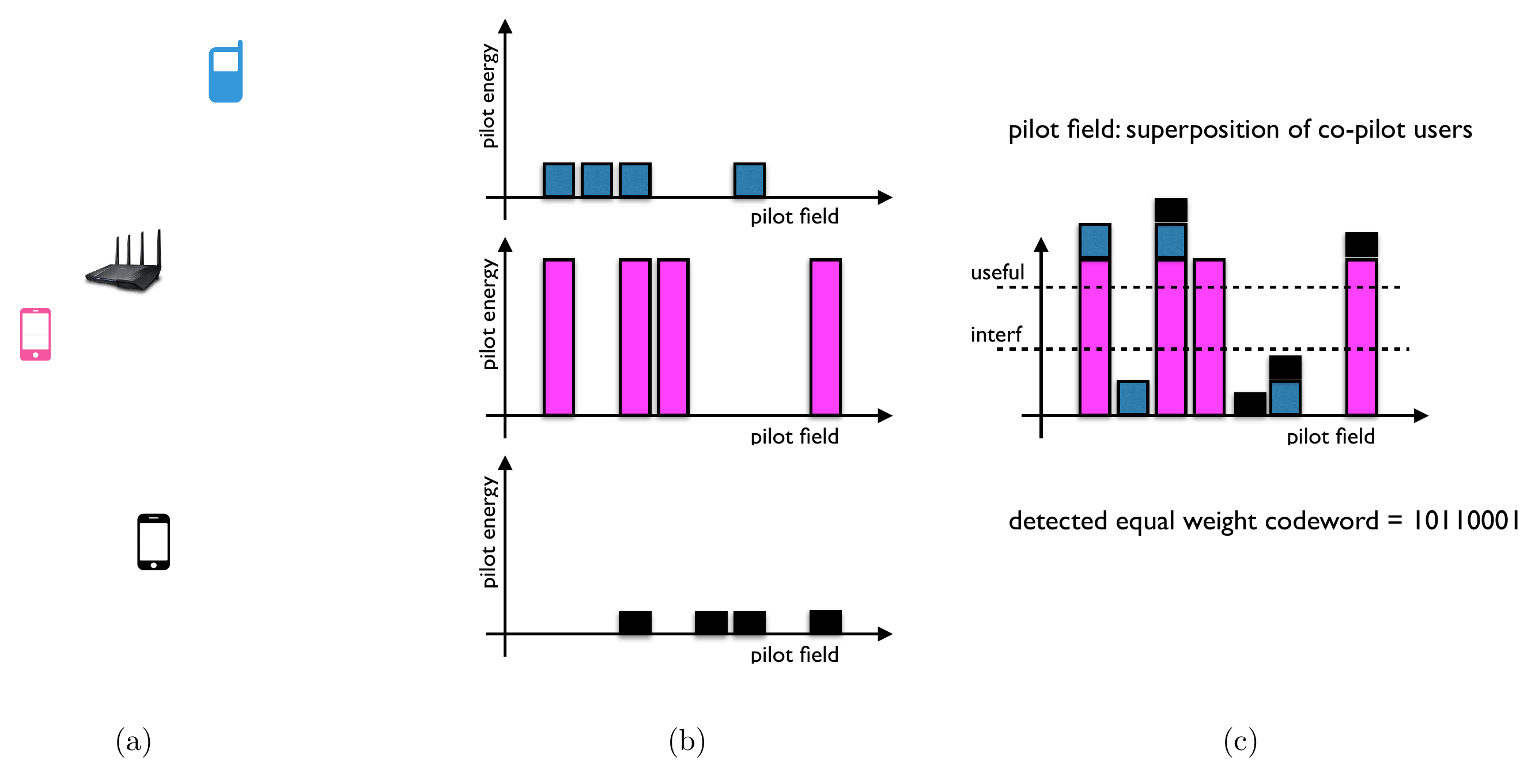}}
\caption{{\small A qualitative example of the coded pilot threshold rule for $Q' = 8$ and three co-pilot users. 
a) geometry with one RRH and three users; b) individual equal-weight pilot fields after spatial matched filtering; c) received superposition and two-threshold 
mechanism. In this case, the pilot $q$ can be trusted.}}
\label{q1q2}
\end{figure}

Fig.~\ref{q1q2} shows qualitatively the threshold rule for a concrete example with $Q' = 8$, where the sum in (\ref{qpilot2}) contains one strong user
and two weaker co-pilot users. In general, a pilot $q$ can be trusted even in the presence of a large number of users in the same group $q$, 
provided that their collective sum power is indeed sufficiently weak. 

\subsection{System Operation and Achievable Spectral Efficiency}  \label{sec:Fog operation and signal expr}

For a sufficiently dense deployment of RRHs, each UE can be potentially served 
by several RRHs. The network layer routing ensures that the RRHs in the vicinity of any given UE $j$ have 
data packets in their queue, ready for transmission to user $j$. 
After receiving the UL pilot of the current TDD slot, RRH $k$ forms a list $\Uc_k$ of users (referred to as the {\em active set} of RRH $k$)
that can be served in the corresponding DL data subslot. In particular, 
a user $j : u_j \in \Phiuq$ is in the active set if: 
1) RRH $k$ determines (by applying the aforementioned trusted pilot detection rule) that the corresponding pilot $q$ is trusted. Since an equal-weight codeword is uniquely assigned to the users in pilot group $\Cc_q$, the corresponding strong user $j : u_j \in \Phiuq$ in the proximity of RRH $k$ is also identified; 2) RRH $k$ has a DL data packet ready to send to user $j$.  

By construction, we have $0 \leq |\Uc_k|  \leq Q$ and we assume that the number of RRH antennas is $M \geq Q$, 
in compliance with the massive MIMO concept of more antennas than users \cite{Marzetta-massiveMIMO}.
The $k$-th RRH calculates the precoding vectors $\bv_{k,j}$ for each user $j \in \Uc_k$ from its (trusted) channel estimates. 
We introduce the notation $\widehat{\gv}_{k,j} = \widehat{\gv}_k^{(q)}$ where user $j$ is the unique user in $\Uc_k$ associated to pilot group $q$.  
The precoder is obtained by using zero-forcing beamforming (ZFBF) as in \cite{massiveMIMO-book-Marzetta}. Here, however, we consider unit-norm precoding vectors. These are obtained by first arranging the estimated channel vectors into the 
$M \times |\Uc_k|$ channel matrix $\widehat{\Gm}_k = \{ \widehat{\gv}_{k,j} : j \in \Uc_k\}$, 
then computing the Moore-Penrose pseudo-inverse matrix $\widehat{\Gm}_k^{\dagger} = \widehat{\Gm}_k (\widehat{\Gm}_k^\herm \widehat{\Gm}_k)^{-1}$, 
and finally letting $\bv_{k,j}$ be the column of $\widehat{\Gm}_k^{\dagger}$ corresponding to $\widehat{\gv}_{k,j}$, normalized by its magnitude.

Notice that a given user $j : u_j \in \Phiuq$ may be in the active set of multiple RRHs. 
Since the RRHs make their own local decisions without coordination (apart from the  implicit coordination induced by network layer routing), each RRH whose active set contains user $j$ will send
simultaneously the same data packet to user $j$, achieving implicit macro-diversity.   We let $\Ac_j = \{k : a_k \in \Phib, j \in \Uc_k\}$ denote the set of RRHs 
serving user $j$, and denote by $\Ac_j^c$ the complement set with respect to $\Phib$. 
The DL received signal at user $j$ is given by 
\begin{align} \label{DL}
y_j &=  \left ( \sum_{k \in \Ac_j}   \gv_{k,j}^\herm \bv_{k,j} \right ) x_{j}  + \sum_{k \in \Ac_j}  \gv_{k,j}^\herm \left ( \sum_{j' \in \Uc_k : j' \neq j}  \bv_{k,j'} x_{j'} \right ) +  \sum_{k \in \Ac^c_j}  \gv_{k,j}^\herm \left ( \sum_{j' \in \Uc_k}  \bv_{k,j'} x_{j'} \right ) + z_j  
\end{align}
where the first term is the useful signal, the second and third terms correspond to the interference from RRHs serving user $j$ and RRHs not serving user $j$, respectively, while $z_k \sim \Cc\Nc(0,\sigmaNtwo/\Pstrfog)$ represents the AWGN with $\Pstrfog$ the per-stream DL power and $\sigmaNtwo$ the noise variance.
We assume unit-energy, zero-mean, and mutually statistically independent data symbols. Hence, the total transmit power
of RRH $k$ in the current DL subslot is given by $|\Uc_k| \Pstrfog$. 
When the RRH has no trusted pilots, it simply does not transmit anything thus achieving power savings in a seamless way (automatic shut-down in the absence of 
requested traffic).  For future comparison with the baseline cellular system,  the per-RRH 
average transmit power is given by $\EE[|\Uc_k|]\Pstrfog$.

We consider the ergodic spectral efficiency where ergodicity is with respect to the small scale fading, 
for fixed realization of $\Phiu$ (user positions) and $\Phib$ (RRH positions). 
We start from the general achievable ergodic spectral efficiency expression for user $j$ \cite{massiveMIMO-book-Marzetta}\footnote{This is generally referred to as a 
``lower bound'' but we remark here that a lower bound to an achievable spectral efficiency is achievable, therefore, we will simple treat it as our achievable spectral efficiency benchmark.}
\begin{equation} 
C_j = \log_2 \left ( 1 + \frac{\left | \EE[ \mbox{useful} ] \right |^2}{ \frac{\sigmaNtwo}{\Pstrfog} + {\rm Var}(\mbox{useful}) + \EE[ | \mbox{interference} |^2] } \right ) 
\label{rate1}
\end{equation}
and derive a compact expression that will be used in the next section for the stochastic geometry analysis.  
Before going into the ergodic rate analysis, it is convenient to distinguish between the contribution of non-copilot interference and that of co-pilot interference. To this purpose, we further partition the set of interfering RRHs $\Ac_j^{\rm c}$ into a set $\widetilde{\Ac}_j^{\rm c}$ of RRHs for which pilot $q$ is trusted, and a set 
$\overline{\Ac}_j^{\rm c}$ of RRHs for which pilot $q$ is untrusted. 
The non-copilot interference is due to all data streams transmitted by RRHs $k \in \overline{\Ac}_j^{\rm c}$ 
as well as all data streams from RRHs $k \in \Ac_j \cup \widetilde{\Ac}_j^{\rm c}$ except those associated to the 
channel estimate obtained from UL pilot $q$.  The exact closed-form expression for an  achievable ergodic rate in (\ref{rate1}) is given by:

\begin{proposition}  \label{prop:Erg rate Fog}
For the fog massive MIMO network model defined above,  the following ergodic spectral efficiency is achievable by ZFBF:   
\begin{align} 
C_j = \log_2 \left ( 1 + \frac{\left | \sum\limits_{k \in \Ac_j}  \sqrt{\alpha_{k,j} \beta_{k,j}} \sqrt{M - |\Uc_k|+1} \right |^2}
{ \frac{\sigmaNtwo}{\Pstrfog}  - \sum\limits_{k \in \Ac_j \cup \widetilde{\Ac}_j^{\rm c}} \alpha_{k,j} \beta_{k,j} |\Uc_k|
	+ \sum\limits_{k: \ak  \in \Phib}  \beta_{k,j} |\Uc_k| + \sum\limits_{k \in \widetilde{\Ac}_j^{\rm c}} \alpha_{k,j} \beta_{k,j} (M - |\Uc_k|+1)} \right ), 
\label{rate-ZFBF}
\end{align}
where 
\begin{equation} \label{mmse-scaling}
\alpha_{k,j} = \frac{\beta_{k,j}}{\sum_{j' : u_{j'} \in  \Phiu^{(q)}} \beta_{k,j'}   + \frac{\sigmaNtwo}{\Pu Q} }
\end{equation}
is the scaling coefficient in the conditional mean $\EE[ \gv_{k,j} | \widehat{\gv}_k^{(q)}] = \alpha_{k,j} \widehat{\gv}_k^{(q)}$, 
and where $\widehat{\gv}_k^{(q)}$  given in (\ref{qpilot1}).\footnote{Notice that since $ \gv_{k,j}$ and $\widehat{\gv}_k^{(q)}$ are jointly Gaussian with zero mean and IID components with respect to the space (antenna)  dimension,  the conditional mean coincides with the linear MMSE estimator of $\gv_{k,j}$  given $\widehat{\gv}_k^{(q)}$, which reduces to a scaling by the factor $\alpha_{k,j}$.}

\begin{proof}
	Appendix~\ref{Appendix:proof prop erg rate fog}.
\end{proof}

\end{proposition}

It is important to notice that the MMSE scaling factors $\alpha_{k,j}$ appear in (\ref{rate-ZFBF}) uniquely as a product of the analysis, but have nothing to do with MMSE channel estimation. In fact, the channel estimates from the UL pilots are given by (\ref{qpilot1}), which is simply a Least-Squares 
projection  that does not assume any prior knowledge of the large-scale channel gains. 

In order to obtain a simple expression, amenable to the stochastic geometry analysis of the next sections, we consider the limit 
$M \rightarrow \infty$ as in \cite{Marzetta-massiveMIMO}.
Notice that the coefficients $\alpha_{k,j}$ defined in (\ref{mmse-scaling}) include the path coefficients of  all UEs $j' : u_{j'} \in  \Phiuqact$.  
These dependencies make the stochastic geometry analysis intractable. However, we notice that the denominator of the fraction in the RHS of (12)  
contains a single strong channel gain and many other weak channel gains, 
as a  consequence of the trusted pilot detection rule.  
By the symmetry of the PPP, the mean value of this denominator is independent 
of the RRH index $k \in \Ac_j \cup \widetilde{\Ac}_j^{\rm c}$. Furthermore, these denominators should be close to 
their mean value and therefore, up to small statistical fluctuations, all approximately equal. With this approximation, 
letting $M \rightarrow \infty$ in (\ref{rate-ZFBF}), we obtain the large-$M$ spectral  efficiency expression
\begin{equation} 
\Cjinfty= \log \left ( 1 + \frac{\left | \sum_{k \in \Ac_j}  \beta_{k,j} \right |^2}
{\sum_{k \in \widetilde{\Ac}_j^{\rm c}} \beta_{k,j}^2} \right ),  
\label{rate-inf}
\end{equation}
which is well-known from \cite{Marzetta-massiveMIMO}, with the only difference of the 
coherent combining of multiple transmissions in the numerator in the case $|\Ac_j| \geq 1$ for the already discussed inherent 
macro-diversity of the scheme.

\section{Main Analytical Results} \label{sec:pilot allocation scheme}

First, we present analytical expressions for the active co-pilot user density and the active RRH density. 
Then, we derive the approximate expression for the spatially averaged spectral efficiency in the limit of $M \to \infty$.

\subsection{Active Co-pilot User Density}

Recall from Section~\ref{sec:on-the-fly pilot contam control} that each RRH identifies its trusted pilots based on a decision rule involving 
the two thresholds $\ThreSig$ and $\ThreInt$. Assuming that the large-scale channel gain coefficients 
appearing in \eqref{qpilot2} are decreasing functions of the distance between UE and RRH with a radial symmetry, 
for the sake of the stochastic geometry analysis we shall approximate the trusted pilot detection rule with a purely geometric rule. 
Namely, the thresholds $\ThreSig$ and $\ThreInt$ correspond to two circular regions each RRH referred to in the following as the 
{\em coverage disk} and the {\em protection disk} with radii $\Rin$ and $\Rout = (1 + \epsilon) \cdot \Rin$, with $\epsilon  \geq 0$, respectively. 

Without loss of generality, we can place a reference user at $\uo$, associated to a given  pilot group $\Cc_{q}$, referred to for simplicity as 
``pilot $q$''. The reference user can be served by a RRH at location $a_0$  if an only if 
%
$\ao \in \Bcal_{u_0} (\Rin)$ (the RRH is within the inner radius of the reference user) and $\ao \notin \Bcal_{\uj} ( \Rout)$ for all
$\uj \in \Phiuq \setminus \uo$ (the RRH is outside the protection radius of all other users associated to the same pilot $q$).
We say that a user at $u_0$ is ``allowed'' to receive in the DL if there is at least one RRH at some position $a_0$ meeting the above condition. 
Let $\theta$ denote the fraction of the disk $\Bcal_{u_0} (\Rin)$ uncovered by the union of disks 
$\bigcup_{\uj \in \Phiuq \setminus \uo} \Bcal_{\uj} ( \Rout)$.  Then, conditioned on the {\em uncovered fraction} $\theta$, we have 
\begin{align} \label{eq:allowed prob cond}
\Pbb \left( u_0 \,\textnormal{ is allowed} \, | \theta \right) &= 1 - e^{\lambdab \pi \Rin^2 \theta}
\end{align}
which follows from 
the void probability of the PPP $\Phib$ (the probability that there exists no point of the PPP~\cite{StocGeo-Haenggi}) in the uncovered 
region of area $\pi R_{\rm in}^2 \theta$.  Averaging (\ref{eq:allowed prob cond}) over $\theta$ gives
\begin{align}
\Pbb \left( u_0 \,\textnormal{ is allowed} \right) 
&= \int_{0}^{1} \left( 1 - e^{\lambdab \pi \Rin^2 \theta} \right)  \, f_{\theta} (\theta) \, \Dd\theta  \label{eq:prob u allowed}
\end{align}
where $f_{\theta} (\cdot)$ denotes the probability density function (PDF) of the uncovered area fraction $\theta$.
As a result, the mean active co-pilot user density is given by 
\begin{align}
\lambdauqact &= \lambda \, \Pbb \left( \uo \,\textnormal{ is allowed} \right)  
= \lambda  \int_{0}^{1} \left( 1 - e^{\lambdab \pi \Rin^2 \theta} \right)  \, f_{\theta} (\theta) \, \Dd\theta . \label{eq:co-pilot dens def}
\end{align}

In light of the fact that there are no available tractable expressions for  the PDF $f_\theta(\cdot)$, 
a possible {\em semi-analytic approach} consists of using  Monte-Carlo simulation
to calculate $f_\theta(\cdot)$ for any given ratio of $\lambdab/\lambda$ and evaluate the integral in (\ref{eq:allowed prob cond}).
As an alternative, in order to gain analytical insight, we extend the approximation proposed in~\cite{unique-coverage-Haenggi} to calculate $f_\theta(\cdot)$ to the case
$\Rin \leq \Rout$. The accuracy of this approximation is examined later in the section (cf. Example~\ref{eg:co-pilot density validattion}).
Using this approach, $f_{\theta} (\cdot)$ can be approximated by a mixed-type PDF formed by the convex combination of a uniform PDF, 
a point mass at $\theta = 0$, and a point mass at $\theta = 1$. 
The resulting approximated PDF is given by (cf. Appendix~\ref{proof:PDF of Aun estimate})
\begin{align} \label{eq:PDF of Aun estimate} 
{\hat{f}}_{\theta} (\theta) &=
\begin{cases}
1 +  e^{-\pi \lambda (\Rout + \Rin)^2}  - 2 \, e^{-\pi \lambda \Rout^2}    & \quad \textnormal{point mass at} \,\, \theta = 0  \\
2 \left( e^{-\pi \lambda \Rout^2}  - e^{-\pi \lambda (\Rout + \Rin)^2}  \right)   & \quad 0 < \theta < 1  \\
e^{-\pi \lambda (\Rout + \Rin)^2}   & \quad \textnormal{point mass at} \,\, \theta = 1 
\end{cases}
\end{align}
\hfill $\lozenge$

Using (\ref{eq:PDF of Aun estimate}) in~(\ref{eq:co-pilot dens def}) yields the closed-form approximation for the active co-pilot user density:
\begin{align} \label{eq:copilot user dens est}
\lambdauqact
&\approx 2 \lambda \left(e^{-\pi  \lambda  \Rout^2}-e^{-\pi  \lambda (\Rin+\Rout)^2} \right) \left( 1- \frac{1-e^{-\pi  \lambdab  \Rin^2}}{\pi  \lambdab  \Rin^2}\right)  + \lambda \left(\frac{1-e^{-\pi  \lambdab  \Rin^2}}{e^{ \pi  \lambda  (\Rin+\Rout)^2}}\right) .
\end{align}

\begin{figure}[h]
	\centering
	\includegraphics[scale=0.32]{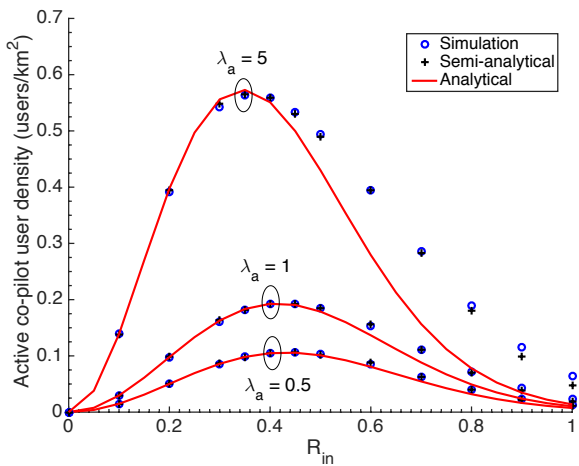}
	\caption{\small Active co-pilot user density v. $\Rin$ for $\lambdab = 0.5$ RRHs per unit area, $\lambdab = 1$ RRH per unit area and $\lambdab = 5$ RRHs per unit area, with $\lambdauq = 1$ user per unit area and $\Rout = 1.25 \Rin$.} 
	\label{Fig:Copilot user dens example1}
\end{figure}

\begin{example} \label{eg:co-pilot density validattion}
Fig.~\ref{Fig:Copilot user dens example1} compares, as function of $\Rin$, the active co-pilot user density $\lambdauqact$ estimated via (\ref{eq:co-pilot dens def}) and (\ref{eq:copilot user dens est}) against the simulated counterpart for user density $\lambdauq = 1 $ user per unit area,  and 
different RRH densities $\lambdab = 0.5$ RRHs per unit area, 
$\lambdab = 1$ RRH per unit area,  and $\lambdab = 5$ RRHs per unit area, and with $\Rout = 1.25 \Rin$.
These settings span the three possible network scenarios of $\lambdab < \lambdauq$, $\lambdab = \lambdauq$ and $\lambdab > \lambdauq$, 
respectively.  As seen in Fig.~\ref{Fig:Copilot user dens example1}, the value of $\Rin$ maximizing the active co-pilot user density decreases 
with $\lambdab$ (when $\lambdauq$ is fixed).
The semi-analytic approach is also shown.  As said before, this approach is very accurate 
and significantly more computationally efficient than the full Monte-Carlo system simulation.
As can be seen in Fig.~\ref{Fig:Copilot user dens example1}, a very satisfactory agreement is observed when $\Rin$ is less than or equal 
to the value that maximizes $\lambdauqact$. If $\Rin$ is beyond the value that maximizes $\lambdauqact$, then (\ref{eq:copilot user dens est}) 
underestimates the active co-pilot user density and this estimation error is noticeable when $\lambdab >1$. 
Similar agreement has been observed for other values of parameters.  
Therefore, it is reasonable to utilize (\ref{eq:copilot user dens est}) as a proxy to optimize the active co-pilot user 
density up to $\Rin$ that maximizes $\lambdauqact$.
\hfill $\lozenge$
\end{example}

\subsection{Active RRH Density and Transmit Power}

The RRH at $\ao$ has an active user corresponding to the pilot under consideration if and only if there exists only one user in its coverage disk, i.e., $\exists 
\uo \in \Bcal_{\ao} (\Rin) \cap \Phiuq$ and no other users in its protection disk, i.e., 
$\Bcal_{\ao}(\Rout) \cap  \Phiuq \setminus \{u_o\} = \emptyset$. 
Denoting by $\Phibact^{(q)}$ the locations of active RRHs corresponding to the pilot 
and by $\lambdabact^{(q)}$ the density of active RRHs, we can express
\begin{align}
	\lambdabact^{(q)} 
	&= \lambdab \, \Pbb \left( \Phiuq( \Bcal_{\ao} ( \Rin) ) = 1 \right)  \, \Pbb \left( \Phiuq( \Bcal_{\ao} (\Rout) \setminus \Bcal_{\ao} (\Rin) ) = 0 \right)  \label{eq:RRH dens active stp 1} \\
	&= \lambdab \, \lambdauq \pi \Rin^2 \, e^{ - \pi \lambdauq \Rout^2 } \label{eq:RRH dens active}
\end{align}
where for a region $B \subseteq \RR^2$ we use the short-hand notation $\Phiuq( B ) = |\Phiuq \cap B|$, and  where (\ref{eq:RRH dens active}) follows 
from the well-known property of PPPs \cite{StocGeo-Haenggi}
$\Pbb \left( \Phiuq( B ) = n \right) = e^{-\lambdauq |B| } \frac{ \left( \lambdauq |B| \right)^n}{ n!}$.

%
%

It is also immediate to express the average number of users served by the RRH as
\begin{align}
\Ebb [ |\Uc_0| ] & = Q  \lambdauq \pi \Rin^2 \, e^{ - \pi \lambdauq \Rout^2 }. 
\end{align}
Multiplying by the per-stream power $\Pstrfog$, we obtain the average RRH transmit power as
\begin{align} \label{eq:Avg TX power Fog}
\Pavg &=  \Pstrfog Q \, \lambdauq \pi \Rin^2 \, e^{ - \pi \lambdauq \Rout^2 } .
\end{align}


\subsection{Spectral Efficiency in the Large Antenna Regime} \label{sec:Seff analysis fog}

Let $\eta$ denote the propagation pathloss exponent and $r_{k,j} = |a_k - u_j|$ the distance between RRH $k$ and UE $j$. 
We assume that the large-scale channel gains obey the usual distance-based polynomial decay law $\beta_{k,j} = r_{k,j}^{-\eta}$. 
Using this in (\ref{rate-inf}), the ergodic spectral  efficiency of the typical user can be written as
\begin{align} 
\Coinfty = \log_2 \left ( 1 + \frac{ \sigmaStwo  }{ \sigmaItwo } \right ),  
\label{rate-inf expanded}
\end{align}
where $\sigmaStwo = \left| \sum_{k \in \Acal_0} r_{k,0}^{-\eta}  \right|^2$ and $\sigmaItwo = \sum_{k \in \widetilde{\Ac}_0^{\rm c} } r_{k,0}^{-2 \, \eta}$.
In turn, (\ref{rate-inf expanded}) involves multiple levels of randomness through $|\Ac_0|$, $r_{k,0} \, \forall k \in \Ac_0$, and $r_{k,0} \, \forall k \in \widetilde{\Ac}_0^{\rm c}$.
We marginalize the spectral efficiency  $\Coinfty$ in multiple steps to characterize the spatially averaged per-user 
spectral efficiency, given by
\begin{align}
\Cbaroinfty
&= \sum_{n=1}^{N} \Pbb \left( |\Acal_0| = n \Big| |\Acal_0|>0 \right) \Ebb  \left[  \Ebb\left[ \log_2 \left( 1 + \frac{\sigmaStwo}{\sigmaItwo}\right) | \sigmaStwo \right] \right] . \label{eq:C user avg one}
\end{align}
The value of $N$ in (\ref{eq:C user avg one}) is chosen sufficiently large such that $\Pbb \left( |\Acal_0| = N+1 \right)$ is negligible, the inner expectation is over the interference (i.e.,  $r_{0,k} \, \forall k \in \widetilde{\Acal}_0^{\rm c}$) while the outer expectation is over the signal (i.e., $r_{0,k} \, \forall k \in {\Ac}_0$). 

The probability that the number of RRHs in the uncovered area equals $n$ can be written as
\begin{align}
\Pbb \left( |\Acal_0| = n \Big| |\Acal_0|>0  \right) 
&= \frac{ \int\limits_{0}^{1}\Pbb \left( \Phib(\theta \Bcal(0,\Rin) ) = n \right) f_{\theta}(\theta) \, \Dd \theta}{ \sum\limits_{n'=1}^{N} \int\limits_{0}^{1} \Pbb \left( \Phib( \theta \Bcal(0,\Rin)  ) = n' \right) f_{\theta}(\theta) \, \Dd \theta  }. \label{eq:prob for n RRHs}
\end{align}
This can be computed either by using the PDF of the uncovered area fraction $f_{\theta}(\theta)$ obtained via simulation (semi-analytic method), or
using approximation (\ref{eq:PDF of Aun estimate}). By invoking (\ref{eq:PDF of Aun estimate}), the integral in (\ref{eq:prob for n RRHs}) is given by 
\begin{align}
 &\!\!\!\!\!\!  \int\limits_{0}^{1}\Pbb \left( \Phib(\theta \Bcal(0,\Rin) ) = n \right) f_{\theta}(\theta) \, \Dd \theta  \nonumber \\
 &= \frac{2}{n! \, \pi \lambdab \Rin^2 } \left( e^{-\pi \lambdauq \Rout^2}  - e^{-\pi \lambdauq (\Rout + \Rin)^2}  \right)  e^{-\pi  \lambdauq (\Rout^2 + (\Rout+\Rin)^2 ) }  \bar{\Gamma} \left(n+1,\pi  \Rin^2 \lambdab \right)  \nonumber \\
 &\quad + e^{-\pi \lambdab \Rin^2  -\pi \lambdauq (\Rout + \Rin)^2 } \frac{ (\pi \lambdab \Rin^2 )^n }{n!} .
\end{align}
where $\bar{\Gamma} (\cdot, \cdot)$ is the lower incomplete Gamma function.

Next, by virtue of \cite[Lemma~1]{useful-Lemma-capacity}, the inner expectation in (\ref{eq:C user avg one}) can be expressed as
\begin{align}
 \Ebb \left[ \log_2 \left( 1 + \frac{ \sigmaStwo }{ \sigmaItwo  } \right) | \sigmaStwo \right] 
&= \Ebb \left[  \int_{0}^{\infty} \frac{1}{\gamma} \left( e^{- \gamma \, \sigmaItwo} -e^{-\gamma (\sigmaItwo + \sigmaStwo)} \right) \, \Dd \gamma \right]  \label{eq:Cuser step1} \\
&=  \int_{0}^{\infty} \frac{1}{\gamma} \left( \Ebb \left[ e^{- \gamma \, \sigmaItwo} \right]  - \Ebb \left[ e^{-\gamma \, \sigmaItwo } \right] e^{-\gamma \, \sigmaStwo } \right) \, \Dd \gamma  \label{eq:Cuser step2} 
\end{align}
Due to the activation of the RRH through the on-the-fly pilot contamination control mechanism 
(cf. Section~\ref{sec:on-the-fly pilot contam control}), the interfering RRH locations in the downlink (i.e., appearing in the set $\widetilde{\Ac}^{\rm c}_0$)
are dependent on their user locations and violate the PPP condition. 
In order to overcome this obstacle, we borrow a modeling assumption that was shown to be tight in different 
scenarios \cite{Bacceli2011, MunZhaLozHea-TWC16} and whose validity for our purposes is examined later in Example~\ref{eg:avg user seff cellfree validate}.

\textbf{Assumption 1:} The RRH locations outside the receiver's protection disk $\Bcal_{\uo} (\Rout)$ belong to another 
independent PPP with matched density $\lambdabact^{(q)}$ (cf. (\ref{eq:RRH dens active})).  \hfill $\lozenge$

Under Assumption 1, the expectation of $e^{- \gamma \, \sigmaItwo}$ 
over this PPP yields 
\begin{align}
\Ebb \left[ e^{- \gamma \, \sigmaItwo} \right] 
&= \Ebb \left[ \exp \left( - \gamma \, \sum_{k \in \widetilde{\Ac}_0^{\rm c}} r_{k,0}^{-2 \eta} \right) \right] \\
&= \exp \left( -2 \pi \lambdabact^{(q)} \int_{\Rout}^{\infty} \left( 1- e^{-\gamma \, r^{- 2  \eta}} \right) r \, \Dd r \right) \label{eq:LT Intf step0} \\
&= \exp \left( \pi \lambdabact^{(q)} \Rout^2 + \frac{\pi \lambdabact^{(q)} \gamma^{\frac{1}{\eta}} }{\eta}  \bar{\Gamma} \left( -\frac{1}{\eta}, \frac{\gamma}{\Rout^{2 \eta}} \right) \right) \label{eq:LT Intf step1}
\end{align}
where (\ref{eq:LT Intf step0}) follows from the PGFL (probability generating functional) of the PPP \cite{StocGeo-Haenggi} and (\ref{eq:LT Intf step1}) 
follows from the change of variable $ \gamma \, r^{- 2  \eta} \to r'$ and then solving the integral in (\ref{eq:LT Intf step0}).

Then, the remaining randomness in $\Coinfty$ is due to $\sigmaStwo$. Conditioned on $|\Acal_0| = n$, the distance from the intended RRH locations to the user can be ordered as $\roo, \roone, \ldots, r_{n-1,0}$, where $r_{n-1,0}$ is the distance corresponding to the $n$th strongest serving RRH. Then, we have
\begin{align}
\sigmaStwo &= \left( \sum_{k=0}^{n-1} r_{k,0}^{-\eta} \right)^2 .
\end{align}
Here, the intended RRHs are uniformly distributed in the uncovered region of $\Bcal_{0}(\Rin)$ that is arbitrarily shaped and hence their distances are not tractable in general. Again, we make a modeling assumption (the accuracy of which is also validated in Example~\ref{eg:avg user seff cellfree validate}) in order to proceed with the closed-form analysis.

\textbf{Assumption 2:} The serving RRHs are uniformly distributed in the whole disk $\Bcal_{0}(\Rin)$ and the distances $\{ r_{k,0} \}_{k=0}^{n-1}$ are replaced with their expected values.   \hfill $\lozenge$

Under Assumption 2, we have 
\begin{align} \label{eq:sigmaStwo def one}
\sigmaStwo &\approx \left( \sum_{k=0}^{n-1} {\left(\Ebb \left[ r_{k,0} \right]\right)}^{-\eta} \right)^2
\end{align}
where the expected value of the distance between the user and the $m$th closest point when $|\Acal_0|=n$ is
\begin{align} \label{eq:expected dist BPP}
\Ebb \left[ r_{m,0} \right] &= \Rin \frac{\Gamma(m+\frac{3}{2}) \Gamma(n+1) }{\Gamma(m+1) \Gamma( n+\frac{3}{2}) } \qquad m= 0, 1, \ldots, n-1
\end{align}
which leads to a deterministic value for $\sigmaStwo$ in (\ref{eq:sigmaStwo def one}) and thus avoids the outer expectation in (\ref{eq:C user avg one}). 
Finally an approximation for $\Cbaroinfty$ can be obtained by plugging (\ref{eq:prob for n RRHs}), (\ref{eq:Cuser step2}), (\ref{eq:LT Intf step1}), (\ref{eq:sigmaStwo def one}) and (\ref{eq:expected dist BPP})  in (\ref{eq:C user avg one}).




\begin{figure}
	\centering
	{
		\begin{minipage}[b]{.45\linewidth}
			{
				\includegraphics[scale=0.25]{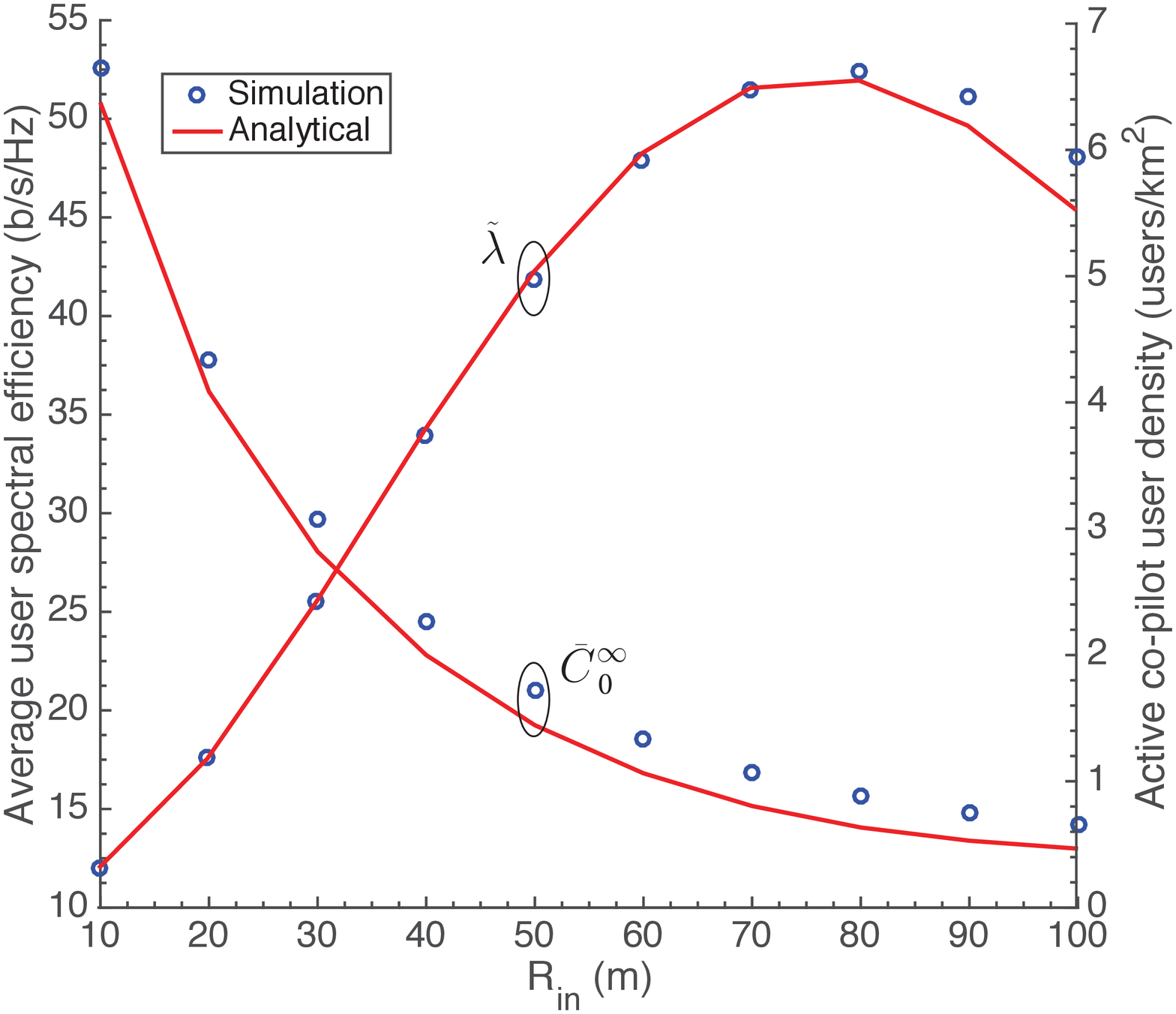} 
			}
			\subcaption{Average user spectral efficiency \label{subfig:Avg user Seff per pilot fog}}
		\end{minipage}
		\begin{minipage}[b]{.45\linewidth}
			{
				\includegraphics[scale=0.25]{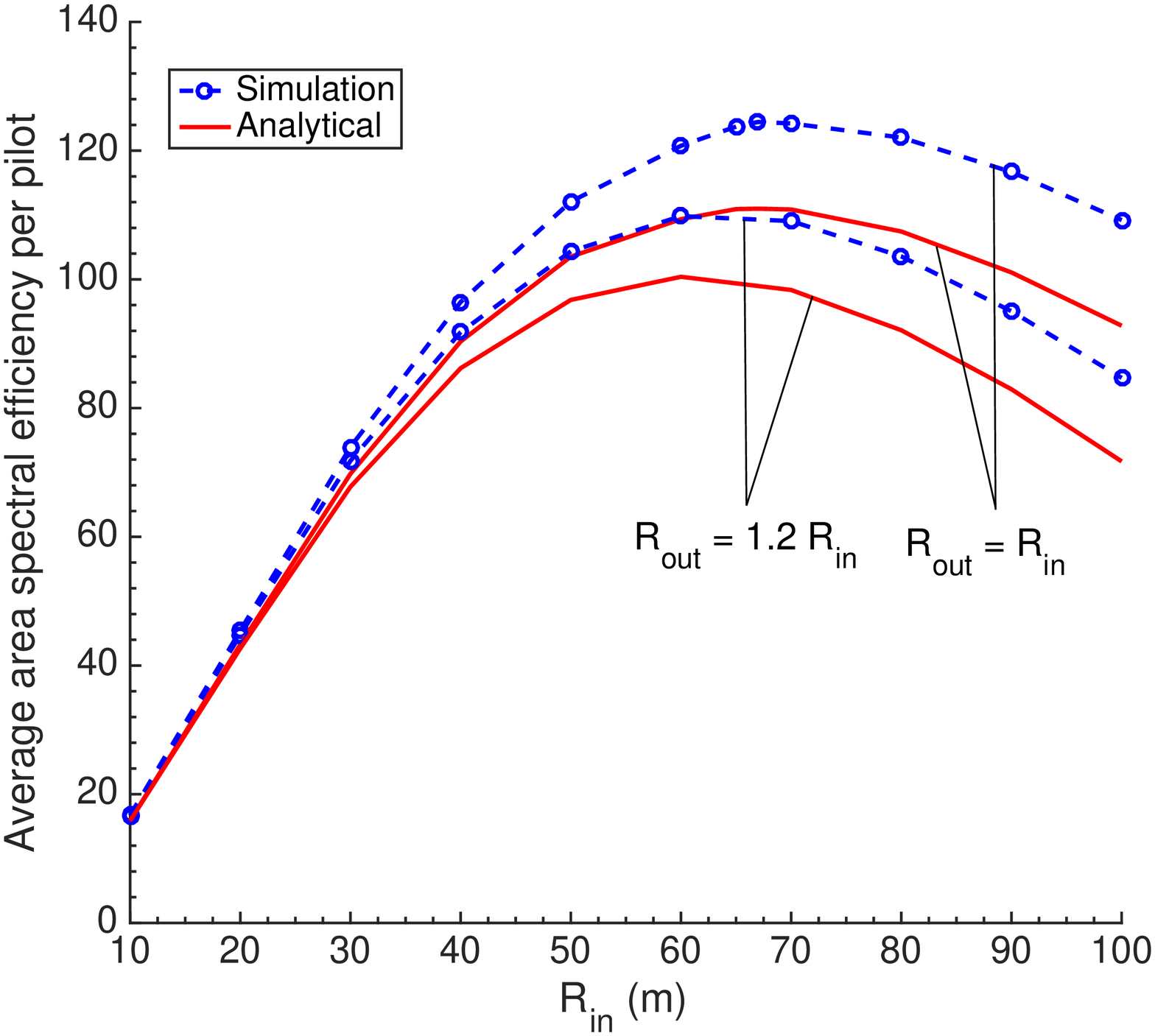} 
			}
			\subcaption{Average area spectral efficiency per pilot \label{subfig:Avg area  Seff per pilot fog}}
		\end{minipage}
		\caption{\small Spectral efficiency v. $\Rin$ for $\lambdab = 31.8 \, \textnormal{RRHs/ km}^2$, $\lambdauq = \lambdab$, $Q=40$ and $\eta=3.75$.}
		\label{Fig:validation seff fog}
	}
\end{figure}

\begin{example} \label{eg:avg user seff cellfree validate}
Consider RRH density $\lambdab = 31.8 \, \textnormal{RRHs/km}^2$ (corresponding to an average of one RRH per circular cell of radius 100 m),  
$\lambdauq = \lambdab$, $\eta = 3.75$, and $Q=40$.
Fig.~\ref{subfig:Avg user Seff per pilot fog} compares the closed-form approximation $\Cbaroinfty$ in (\ref{eq:C user avg one}) based on 
(\ref{eq:PDF of Aun estimate}) and Assumptions 1-2 with its simulated counterpart for protection radius set to $\Rout = 1.2 \Rin$. 
The simulated result corresponds to explicit computation of the spectral efficiency without any modeling assumptions, 
computed through lengthy Monte-Carlo over many network realizations. 
As the figure reveals, the match between analysis and simulation is satisfactory, thereby supporting the validity of the modeling assumptions.\footnote{The validity of these assumptions is also supported by extensive numerical investigations (not reported here, due to space limitations).}
From Fig.~\ref{subfig:Avg user Seff per pilot fog}, we notice that the user spectral efficiency decreases with coverage and protection radii. 
This is because of the increase in the aggregate interference from the active RRHs with $\Rin$ (cf. (\ref{eq:RRH dens active})). 
However, low values of $\Rin$ yield small spatial pilot reuse (cf. Fig.~\ref{subfig:Avg user Seff per pilot fog}). This in turn reduces the spatially averaged area spectral efficiency $(\textnormal{b/s/Hz/km}^2)$ per pilot, defined by $\Cpilot \triangleq \lambdauqact \, \Cbaroinfty$.
Therefore, $\Rin$ must optimized by striking a good tradeoff between  pilot reuse and per-user spectral efficiency. 

Shown in Fig.~\ref{subfig:Avg area  Seff per pilot fog} is $\Cpilot$ with $\lambdauqact$ given by (\ref{eq:copilot user dens est}) and $\Cbaro$ 
given by  (\ref{eq:C user avg one})), alongside  the simulation counterpart. The average area spectral efficiency per pilot $\Cpilot$ 
is uniformly superior with $\Rout=\Rin$ over $\Rout= 1.2 \Rin$.
Again, the match is satisfactory with error due to the modeling assumptions in the range of 5-12\% for the typical values of $\Rin$. 
Therefore, it is reasonable to consider the analytical approximation of $\Cpilot$ as a proxy to tune the radii of the pilot detection rule, while 
the exact value of the area spectral efficiency per pilot can be obtained by conducting simulations 
for specific values of $\Rin$. \hfill $\lozenge$
\end{example}

We conclude by mentioning that, since the signals for different pilots $q$ are completely decoupled 
by the massive MIMO in the regime $M \rightarrow \infty$, the total average area spectral efficiency (b/s/Hz per unit area) of the fog massive MIMO 
system operating with $Q$ pilot groups is simply given by  $\Careafog = Q \Cpilot$. 


\section{Cellular Massive MIMO System} \label{sec: Net model cellular mMIMO}

In this section, we briefly present the system model and achievable spectral efficiency analysis 
for a baseline cellular massive MIMO system that we use as a term of comparison for our fog massive MIMO system. 
Differently from the classical works on massive MIMO (e.g., \cite{Marzetta-massiveMIMO,hoydis2013massive,huh2012achieving,massiveMIMO-book-Marzetta}), 
which have mostly considered regular (e.g., hexagonal) cells and integer pilot reuse factors, 
here we consider  UEs and BSs placed according to the PPPs $\Phiu$ and $\Phib$, respectively, 
and a statistical fractional UL pilot reuse scheme, which is more representative of a ``small-cell'' deployment.  

\subsection{System Configuration}

In  the considered 
cellular massive MIMO system, each user is served by the closest BS, resulting in the standard Voronoi-tessellated network geometry.
As for the fog system, we assume a coherence block of $T$ dimensions, $L$ of which are dedicated to UL mutually orthogonal pilot sequences. 
A reference BS at $\ao$ serves $|\Uc_0| = \min( |\Vcal_0| , \Np)$ users in the DL, 
where $|\Vcal_0|$ denotes the number of users in its Voronoi cell and $0 < \Np \leq L$ is a parameter that controls the (fractional) pilot reuse across the cells. 
Letting $\Np = L$ yields reuse 1 (all pilots are used in all cells, for sufficiently large user density). Otherwise, for $\Np < L$, we assume that $|\Uc_0|$ out of the $L$ available pilots are selected at random and independently across the cells. This implies that the allocation of pilots to users is done without inter-cell coordination. 
Yet, all users served in the same cell use mutually orthogonal pilots.  The per-stream power in cellular system $\Pstrcell$ is fixed such that the average 
power transmitted by each BS is same as the average power of the RRHs of the fog system.
Using the expression for the average number of users served by a BS given in Appendix~\ref{proof: AvgNoUsersPercell}, 
the per-stream power in the cellular system is related to the average BS power $\Pavg$ as
$\Pstrcell = \Pavg /  \Ebb \left[ |\Uc_0| \right]$. 
Furthermore, the location of BSs serving users associate to a specific pilot $q \in [L]$ can be viewed as a thinned version of the original PPP, 
denoted hereafter by $\Phibact^{(q)}$, whose density equals $\pact \lambdab$. The thinning probability $\pact$ equals the probability that 
the pilot is assigned to one of the $\Uc_0$ users, which can be computed as
\begin{align} \label{eq:pact cellular}
\pact 
&=  \frac{ \Np }{L} + \frac{1}{L} \sum_{\ell=0}^{\Np  - 1} (\ell - \Np) \,  \frac{\Gamma(\ell+c)}{ \Gamma(\ell+1) \Gamma(c)} \frac{ (\lambdau)^{\ell} (c\lambdab )^c }{  (c \lambdab +  \lambdau)^{\ell+c} }.
\end{align}

\subsection{Spectral Efficiency in the Large Antenna Regime} \label{sec:cellular SE analysis}

Following the similar approach of fog massive MIMO (cf. Section~\ref{sec:Fog operation and signal expr}), one can derive the ergodic user spectral efficiency of cellular massive MIMO, where in this case $|\Ac_j| = 1$ for any served user at $\uj$. 
For the sake of brevity,  in this section we directly consider the spectral efficiency in the limit of $M \to \infty$ and obtain its spatial averaged quantity. 
The SIR experienced by a typical user located at $\uo$, served  by a BS located at $a_0$,  is expressed as
 \begin{align} 
 \SIR_0 
 &= \frac{ \beta_{0,0}^2  }{  \sum\limits_{k \in \widetilde{\Ac}_0^{\rm c}} \beta_{k,0}^2  } 
 = \frac{ \roo^{-2 \, \eta}  }{ \roone^{-2 \, \eta} + \sum\limits_{k \in \widetilde{\Ac}_0^{\rm c} \setminus \{1\}}  r_{0,k}^{-2 \, \eta}  }  \label{eq:SIR def two cellular} 
 \end{align}
 where in (\ref{eq:SIR def two cellular}) $\roone$ denotes the distance from the user at $\uo$ to the strongest interfering BS 
located at $\aone \in \Phibact^{(q)}$. At this point, as in~\cite{MunMorLoz-TWC15, FrameworkD2D, MunZhaLozHea-TWC16}, we invoke the modeling assumption introduced in Section~\ref{sec:Seff analysis fog}, i.e., the locations of interfering BSs outside $\Bcal_{\uo} (\roone)$ belongs to a PPP with scaled-down density $\pact \lambdab$ and replace the corresponding collective interference term with its spatial average. This yields
\begin{align}
\SIR_0 
&\approx \frac{\roo^{-2 \, \eta} }{\roone^{-2 \, \eta} + \frac{\pi \lambdab \pact}{\eta-1} \roone^{2-2 \eta} } 
= \frac{\deltao^{2  \eta} }{1 + \frac{\pi \lambdab \pact}{\eta-1} \roone^{2} },   \label{eq: user SIR def 3} 
\end{align}
where (\ref{eq: user SIR def 3}) follows by replacing the second term in the denominator of (\ref{eq:SIR def two cellular}) with its 
expectation computed applying  Campbell's theorem
and by introducing $\deltao = \roone/\roo >1$.

Next, the average user spectral efficiency in the cellular massive MIMO network can be written as 
\begin{align} \label{eq:Avg user Seff cellular}
\Cbaroinfty
&= \int_{1}^{\infty} \int_{0}^{\infty}  \log_2 \left( 1 + \frac{ \deltao^{2 \eta}  }{ 1+ \frac{\pi \lambdab \pact}{\eta-1} \roone^{2}  } \right) f_{\roone, \deltao} (\roone, \deltao) \, \Dd \roone \Dd \deltao
\end{align}
where the PDF $f_{\roone, \deltao} (\cdot, \cdot)$ is given by  
\begin{align}
f_{\roone, \deltao} ( \roone, \deltao ) 
&= \pact (2 \pi \lambdab)^2 \left( \frac{\roone}{\deltao} \right)^3 e^{ -\pi \lambdab \roone^2 \left( \pact + \frac{1- \pact}{\deltao^2} \right) }.  \label{eq:PDF of r1a0}
\end{align}
obtained by applying \cite[Lemma 1]{Non-uniform-UE}.
We can obtain the aggregate average spectral efficiency of all the users of a cell by scaling the average user spectral efficiency $\Cbaroinfty$ by the average number of users served by each BS. 
Consequently, the total average area spectral efficiency of the cellular massive MIMO system equals
\begin{align}
\Careacm 
&= \lambdab \Ebb[ |\Uc_0| ] \, \Cbaroinfty . \label{eq:AreaSeff Cellular}
\end{align}
where $\Cbaroinfty$ is given in (\ref{eq:Avg user Seff cellular}) and $\Ebb[ |\Uc_0| ]$ is given in Appendix~\ref{proof: AvgNoUsersPercell}. 

We tested the above approximated analytical expressions against the full system Monte Carlo simulation (cf. Example~\ref{eg:avg seff validattion cellular}), with excellent agreement. 

\begin{table}[h]
	\caption{Average User and Area Spectral Efficiencies of Cellular Massive MIMO}
	\label{tab:Avg user area seff cellular}
	\begin{center}
		\begin{tabular}{|c|c|c||c|c|} \hline
			\multicolumn{1}{|c|}{ $\Np$ } & \multicolumn{2}{|c||}{$\Cbaroinfty$ ($\textnormal{b/s/Hz}$)}&\multicolumn{2}{c|}{$\Careacm$ ($\textnormal{b/s/Hz/km}^2$)} \\
			\cline{2-5}
			 & Analytical    & Simulation  			& Analytical    & Simulation \\ \hline \hline
				10 	 &   10.69    	  &   10.78	       &    2587.9    & 	2613.5		    \\ \hline
				20 	 &   9.73     	  &    9.74 		&    2998.2    &  	3013.5			\\ \hline
				30 	 &   9.62   	  &    9.69		    &    3051.5    &  	3104.5			\\ \hline
				40 	 &   9.60     	  &    9.70         &    3056.9    &  	3118.1			 \\ \hline
		\end{tabular}
	\end{center}
\end{table}

\begin{example} \label{eg:avg seff validattion cellular}
	Consider $\lambdab= 31.8 \, \textnormal{RRHs/km}^2$, $\lambdau= 10 \lambdab$, $L= 40$ and $\eta=3.75$, yielding $\pact = 0.25$.	
	Shown in Table. \ref{tab:Avg user area seff cellular} is a comparison of $\Cbaroinfty$ (cf. (\ref{eq:Avg user Seff cellular})) and $\Careacm$ (cf. (\ref{eq:AreaSeff Cellular})) against their simulated counterparts. As can be seen, the area spectral efficiency increases with $\Np$, which indicate that it is beneficial to utilize all the available orthogonal pilots to serve the maximum number of users (for the RRH and user densities chosen in this example).
\end{example}

\section{Performance Evaluation} 

With the theoretical framework developed in the previous sections, we now proceed to evaluate the performance of optimized fog massive MIMO system and contrast it with the cellular massive MIMO baseline. 
In fog massive MIMO, we assume $\epsilon=0$ (i.e., $\Rout = \Rin$) and pilot dimension $L=60$ where the UL pilot codebook is formed as described in Section \ref{sec:coded UL pilots},
with $Q = 40$ and $Q' = 20$. In cellular massive MIMO, we consider $L = 60$ orthogonal pilot sequences of dimension $60$ and $\Np=L$, which maximizes the area spectral efficiency while minimizing the outage probability.
We choose $\lambdab = 31.8 \, \textnormal{RRHs/ km}^2$, $\eta=3.75$ and $\lambdau = Q \lambda$ for both the  fog and cellular massive MIMO systems. 
In all the following results, the pilot overhead is not taken into account. 
Depending on the number of signal dimensions available per slot $T$, the spectral efficiencies of fog and cellular massive MIMO systems 
should  be scaled by the pilot overhead factor $(1-\frac{L}{T})$.


%

\begin{figure}
	\centering
	{
		\begin{minipage}[b]{.47\linewidth}
			\centering
			{
				\includegraphics[scale=0.25]{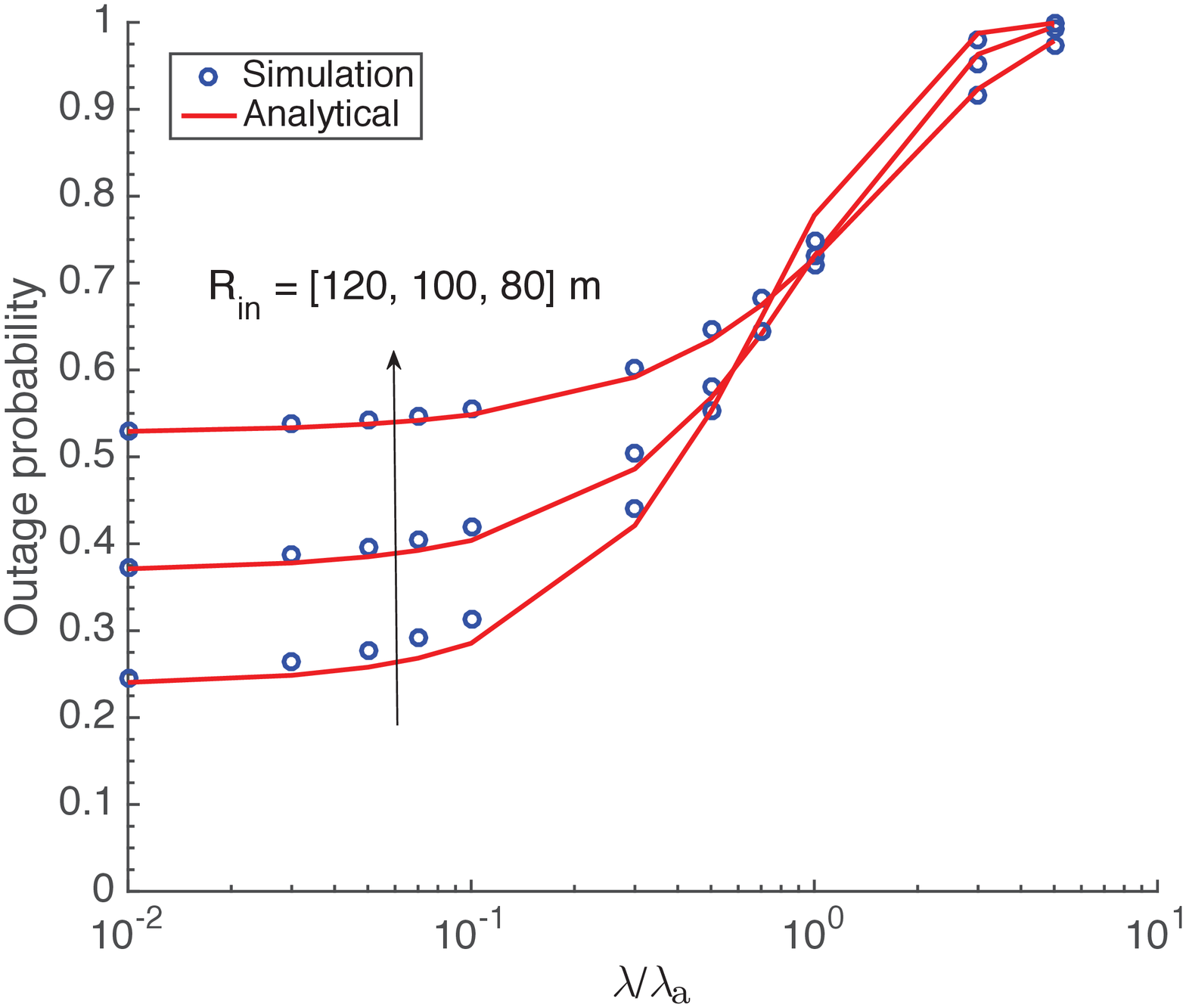} 
			}
			\subcaption{Outage probability \label{subfig:Outageprob v. dens ratio fix Rin}}
		\end{minipage}
		\begin{minipage}[b]{.47\linewidth}
			\centering
			{
				\includegraphics[scale=0.25]{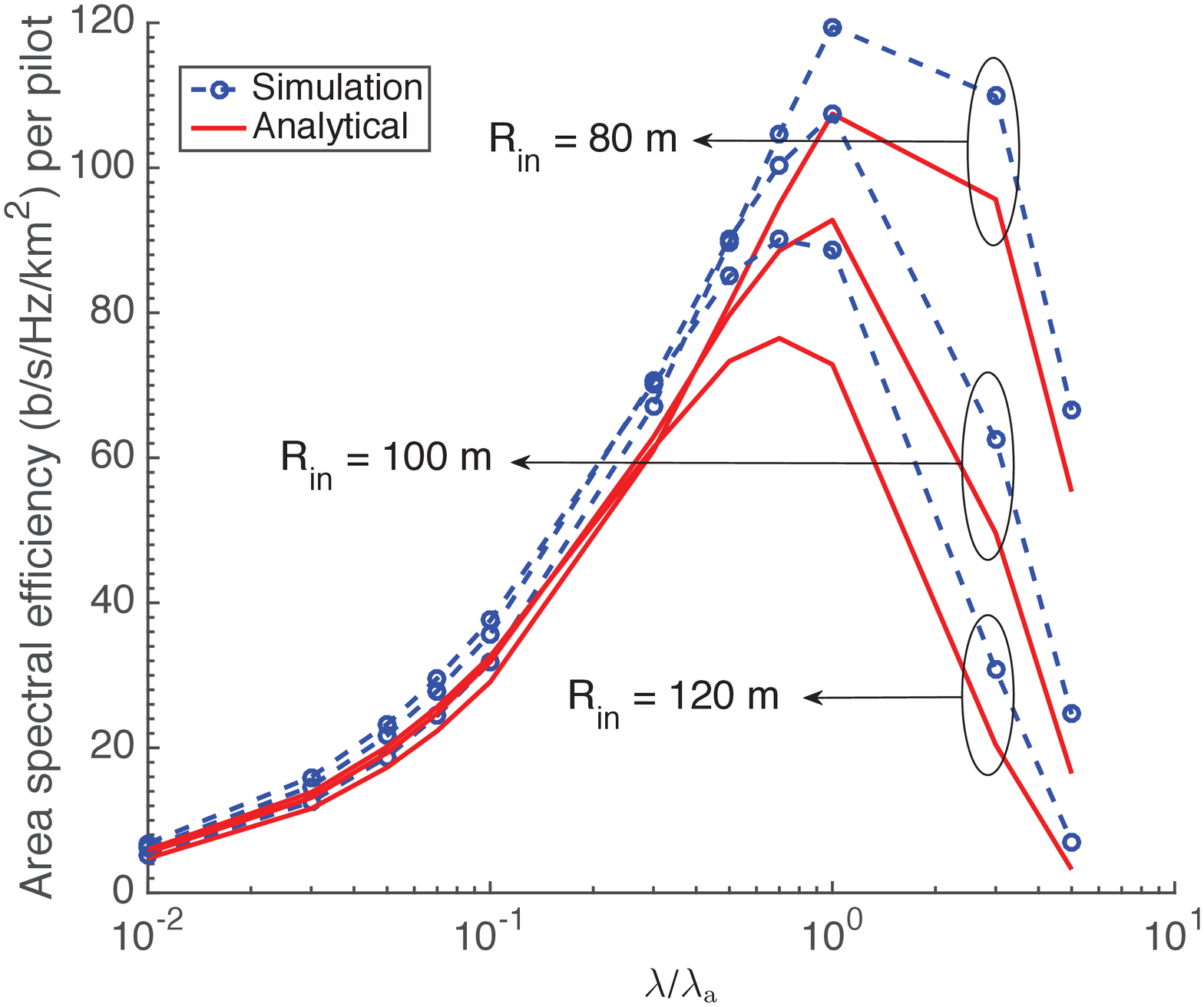} 
			}
			\subcaption{Average area spectral efficiency per pilot \label{subfig:AvgAreaSeff per pilot v. dens ratio fix Rin}}
		\end{minipage}
		\caption{\small Performance of fog massive MIMO as function of $\lambdauq/\lambdab$  with $\Rin = 80$ m, $\Rin = 100$ m and $\Rin=120$ m, $\epsilon =0$, $\lambdab = 31.8 \, \textnormal{RRHs/ km}^2$, and $\eta=3.75$.}
		\label{Fig:performance fog MIMO}
	}
\end{figure}

\begin{example}	\label{eg:Outprob var dens ratio fix Rn}
Fig.~\ref{subfig:Outageprob v. dens ratio fix Rin} shows, as function of $\lambdauq/\lambdab$, the outage probability in a fog massive MIMO network, which is defined as the probability of the event that there is no serving RRH in within $R_{\rm in}$ centered around the user, i.e., 
$P_{\rm out} = \PP( \Bc(0,R_{\rm in}) \cap \Phib = \emptyset)$, 
while Fig.~\ref{subfig:AvgAreaSeff per pilot v. dens ratio fix Rin} shows the corresponding average area spectral 
efficiency per pilot. 
We compare different choices $\Rin = 80$ m, $\Rin = 100$ m and $\Rin=120$ m of the coverage radius. 
As seen in Fig.~\ref{subfig:Outageprob v. dens ratio fix Rin}, the outage probability decreases 
with $\Rin$ at low values of $\lambdauq/\lambdab$, while it increases with $\Rin$ at high values of $\lambdauq/\lambdab$. 
As anticipated before, for a given value of $\Rin$, the area spectral efficiency per pilot (cf. Fig.~\ref{subfig:AvgAreaSeff per pilot v. dens ratio fix Rin}) progressively 
increases with $\lambdauq/\lambdab$ and eventually decreases at higher values of $\lambdauq/\lambdab$ due to the excessive number of pilot collisions, 
essentially preventing most users to be served by any RRH. \hfill $\lozenge$
\end{example}



\begin{figure}
		\centering
	{
		\begin{minipage}[b]{.47\linewidth}
			\centering
			{
				\includegraphics[scale=0.25]{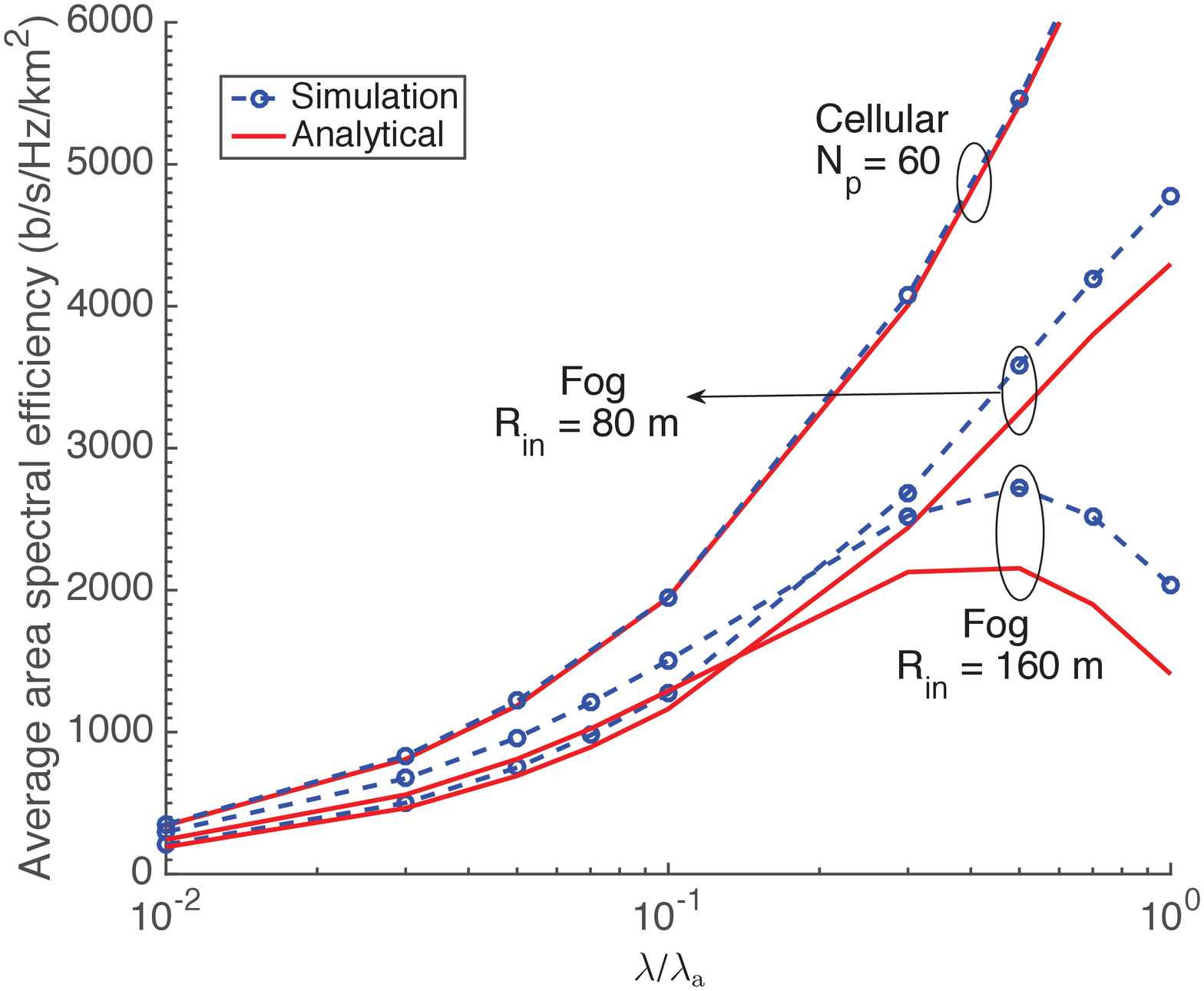} 
			}
			\subcaption{Average area spectral efficiency \label{subfig:AvgAreaSeff Fog cellular infinite}}
		\end{minipage}
		\begin{minipage}[b]{.47\linewidth}
			\centering
			{
				\includegraphics[scale=0.25]{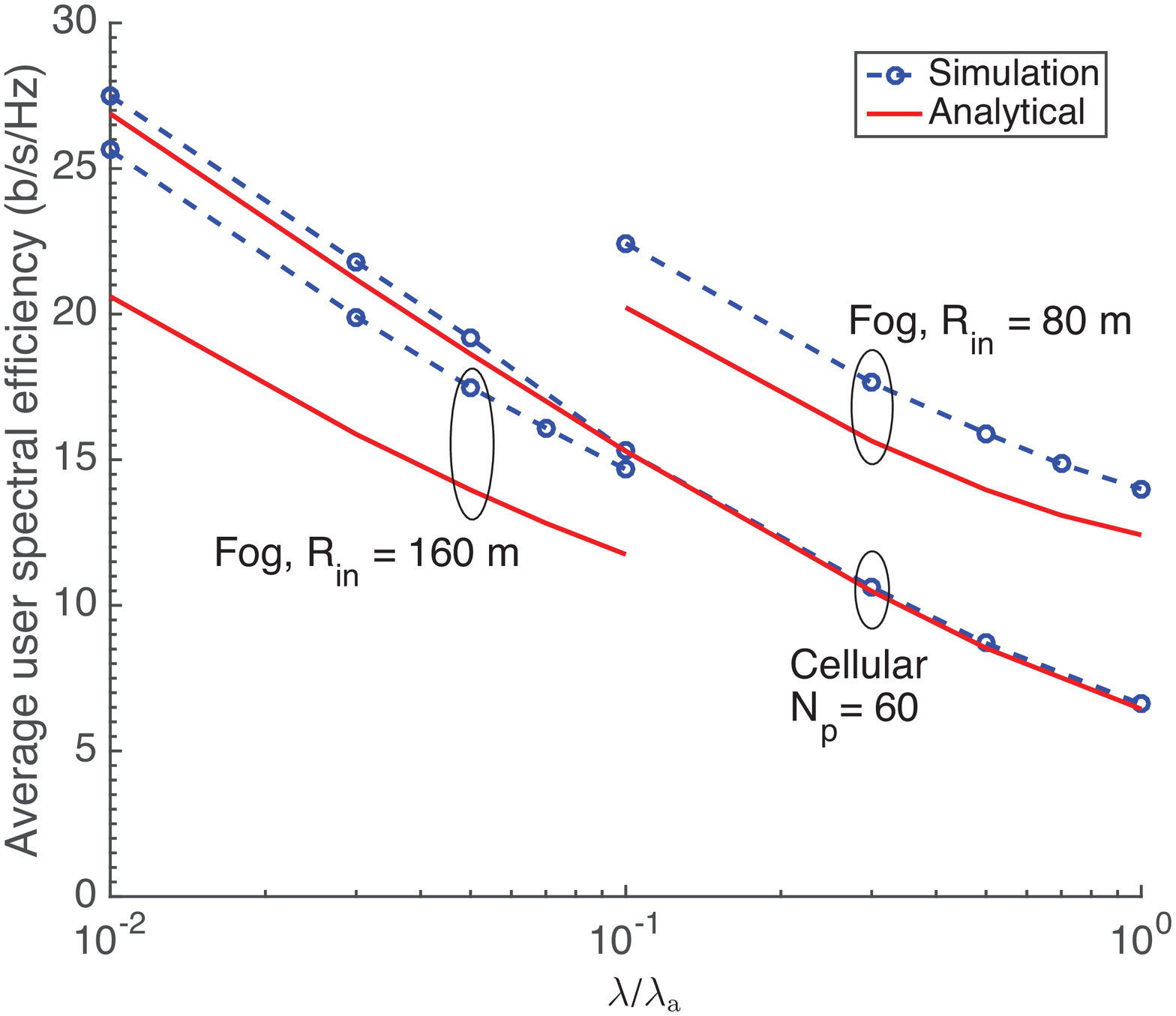} 
			}
			\subcaption{Average user spectral efficiency \label{subfig:AvgUserSeff Fog cellular}}
		\end{minipage}
		\caption{\small Performance evaluation of fog and cellular massive MIMO as function of $\lambdauq/\lambdab$ with $M \to \infty$, $\lambdab = 31.8 \, \textnormal{RRHs/ km}^2$, $\lambdau = Q \lambda$, $Q=40$, $Q'=20$, $L=60$, $\eta=3.75$, $\epsilon=0$ and $\Np=60$.}
		\label{Fig:performance fog MIMO v cellular MIMO}
	}
\end{figure}

\begin{figure}
		\centering
	{
		\begin{minipage}[b]{.47\linewidth}
			\centering
			{
				\includegraphics[scale=0.25]{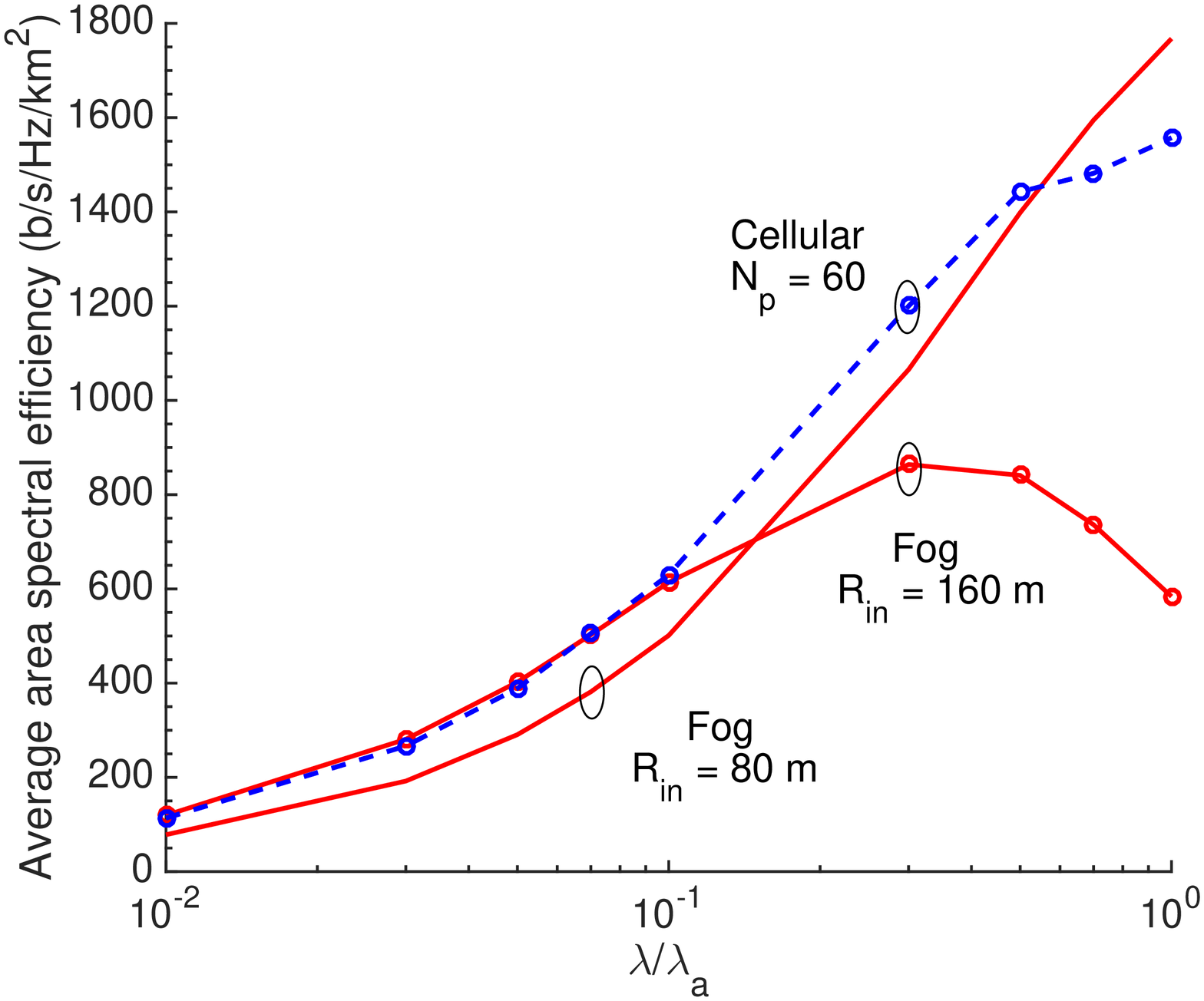} 
			}
			\subcaption{Average area spectral efficiency \label{subfig:AvgAreaSeff Fog cellular M finite}}
		\end{minipage}
		\begin{minipage}[b]{.47\linewidth}
			\centering
			{
				\includegraphics[scale=0.25]{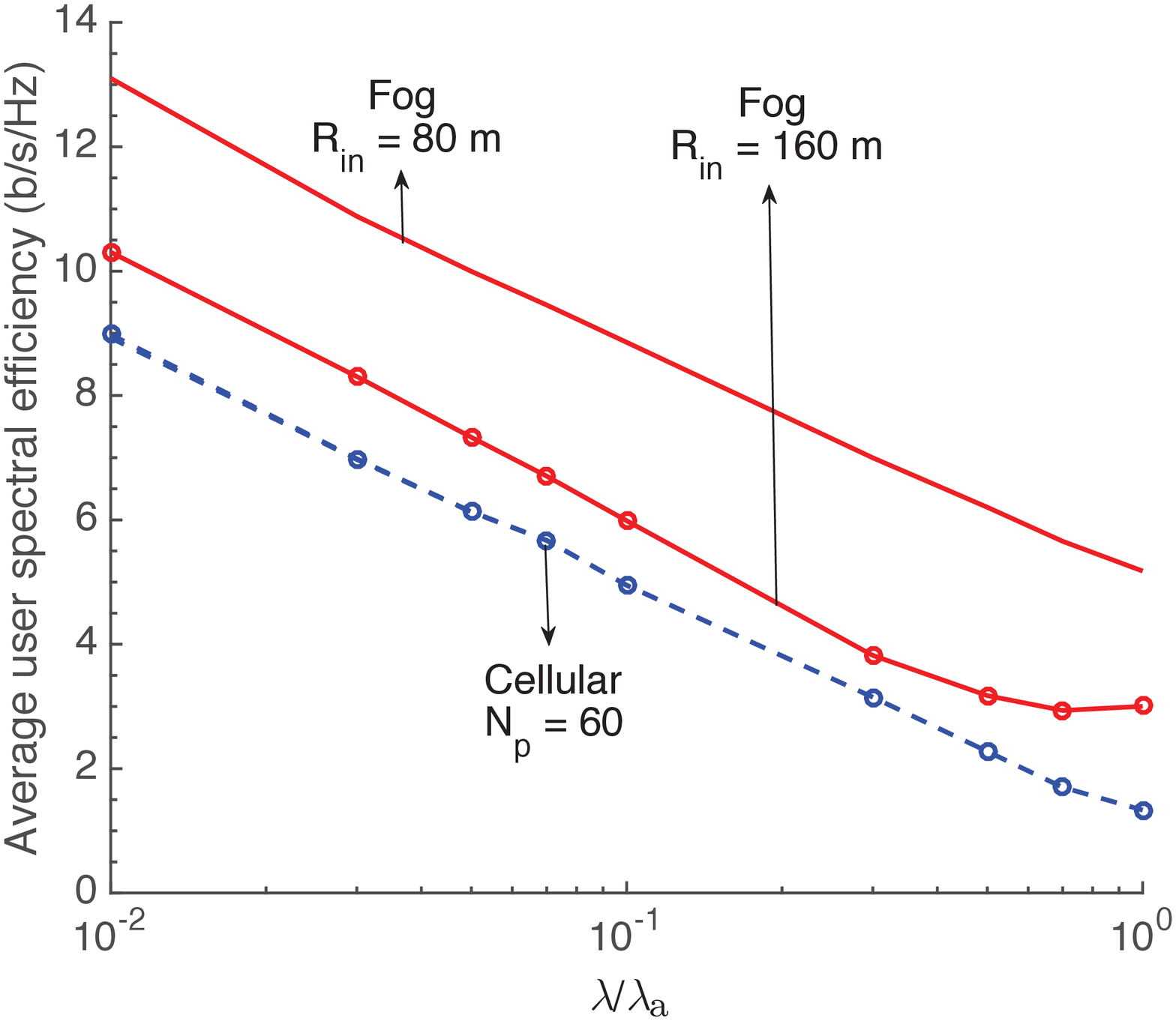} 
			}
			\subcaption{Average user spectral efficiency \label{subfig:AvgUserSeff Fog cellular M finite}}
		\end{minipage}
		\caption{\small Performance evaluation of fog and cellular massive MIMO as function of $\lambdauq/\lambdab$ with $M=64$, $\lambdab = 31.8 \, \textnormal{RRHs/ km}^2$, $\lambdau = Q \lambda$, $Q=40$, $Q'=20$, $L=60$, $\eta=3.75$, $\epsilon=0$ and $\Np=60$.}
		\label{Fig:performance fog MIMO v cellular MIMO finite dim}
	}
\end{figure}

\begin{example}	\label{eg:Fog vs Cellular AreaSeff lightload}
Fig.~\ref{subfig:AvgAreaSeff Fog cellular infinite} compares, as function of $\lambdauq/\lambdab$, the average area spectral efficiencies of fog and cellular massive MIMO systems. The corresponding user spectral efficiencies (restricted to users that are effectively served, i.e., not in outage) 
are shown in Fig.~\ref{subfig:AvgUserSeff Fog cellular}.
For large coverage radius ($\Rin= 160$m),  the outage probability is around $10-20$\% for $\lambdauq/\lambdab \in [0.01, 0.1]$, and the resulting average 
area spectral efficiency is comparable ($15-25$\% lower) to its cellular counterpart. 
A smaller coverage radius ($\Rin=80$ m) outperforms the larger radius for higher user density $\lambdauq/\lambdab \geq 0.3$. \hfill $\lozenge$
\end{example}


\begin{example}	\label{eg:Fog vs Cellular AreaSeff lightload-finiteM}
It is also useful to have an idea of  how meaningful are the analytic results based on stochastic geometry and 
$M \to \infty$ in terms of the qualitative behavior of the system, for the sake of obtaining system design guidelines based on the analysis. 
To this purpose, we performed Monte-Carlo simulations evaluating directly the 
exact finite $M$ ergodic achievable spectral efficiency expression in Proposition \ref{prop:Erg rate Fog} (as well as its counterpart for the cellular case).
As seen in Fig.~\ref{subfig:AvgAreaSeff Fog cellular M finite}, with $\Rin = 160$ m, the area spectral efficiency of fog massive MIMO approaches the value of cellular massive MIMO for $\lambdauq/\lambdab \in [0.01, 0.1]$, and with $\Rin= 80$ m it is slightly higher than its cellular counterpart for  $\lambdauq/\lambdab \geq 0.3$. 
As anticipated before, the average user spectral efficiency of fog massive MIMO (cf. Fig.~\ref{subfig:AvgUserSeff Fog cellular M finite}) 
is uniformly superior over its cellular counterpart when adapting the radius to the user density (in this case, changing from 
$\Rin = 160$ m to $\Rin = 80$ m as $\lambdauq/\lambdab$ increases, which in practice means to decrease the UL pilot power 
as the user density increases). \hfill $\lozenge$
\end{example}

\section{Summary} 

In this paper we have proposed and analyzed an architecture nicknamed ``fog massive MIMO'', where 
a large number of multiantenna RRHs are densely deployed, and serve the UEs using ZFBF, based on channel estimates obtained
from the UL pilots (exploiting TDD reciprocity). 
The key difference between the proposed system and a standard small-cell system is that there is no need for explicit user-cell association, 
and the users are not controlled by the RRHs. In fact, as soon as RRH receives an UL pilot, it can immediately ``beamform back'' a DL data packet
to the user that has sent that pilot, provided that such pilot is not (severely) contaminated.
This ``on-the-fly'' pilot contamination control is obtained using a novel coded pilot approach, that allows 
very simple identification of the uncontaminated pilots at each RRH. The proposed system is completely user-centric, and achieves the following 
advantages over a standard small-cell system: i) RRHs spend transmit power only when they receive uncontaminated pilots from users in their vicinity, 
otherwise they can stay idle and save power; ii) the latency between an UL data request and a DL data transmission is a single TDD resource block (e.g., 1ms); iii) 
the system creates naturally ``hot-spots'', by tuning down the UL pilot power when the user density increases, such that the utilization of
the RRHs is maintained at its optimal value; iv) the users effectively served by the system (i.e., not in outage), achieve a very high rate, much higher than
the cellular counterpart; v) the user-centric operations require no handoff and orthogonal pilot (re)allocation whenever users migrate across the ``fog'' of RRHs, 
thus much lower protocol overhead with respect to conventional small-cell systems especially in the regime of high mobility. 
On the other hand, the proposed system suffers from a relatively large outage probability (fraction of users that are not served by any RRH).
This indicates that such system is suited as a second tier, underneath a conventional cellular tier-1 that provides coverage and connectivity to all users, 
albeit at lower data rates.  
We would like to remark that the investigations in \cite{on-the-fly-ICC16, coded-pilots, sectorized-Zheda}, which consider scenarios with more realistic channel models and where sectorization is exploited, suggest that the proposed systems can outperform the cellular baseline in terms of area spectral efficiency, 
while still preserving the aforementioned low-latency low-complexity user-centric operation.

\appendices
\section{Proof of Proposition~\ref{prop:Erg rate Fog}} \label{Appendix:proof prop erg rate fog}

In order to carry out the ergodic rate analysis, we shall use repeatedly the so-called MMSE decomposition
\begin{equation} \label{mmse-deco}
\gv_{k,j} = \alpha_{k,j} \widehat{\gv}_k^{(q)}  + \ev_{k,j}, 
\end{equation}
where the estimation error vector $\ev_{k,j} = \gv_{k,j} - \alpha_{k,j} \widehat{\gv}_k^{(q)}$ is Gaussian IID 
with mean zero and  per-component variance $\beta_{k,j} ( 1 - \alpha_{k,j})$ and
where $\widehat{\gv}_k^{(q)}$ and $\ev_{k,j}$ are uncorrelated and, because of joint Gaussianity, 
mutually independent.

Let $j$ indicate the reference user for which we wish to compute the useful signal and interference contributions.  
Consider  any RRH-UE pair $(k',j')$ for which the channel estimate $\widehat{\gv}_{k',j'}$ contains $\gv_{k',j}$ (notice: this can happen only if either $j' = j$ or $j' \neq j$ but both users $j$ and $j'$ are associated to  the same pilot group $q$), then 
\begin{eqnarray} \label{ZFBF-coherent}
\gv_{k',j}^\herm \bv_{k',j'} 
& = & \alpha_{k',j} (\widehat{\gv}_{k'}^{(q)})^\herm \bv_{k',j'} + \ev_{k',j}^\herm  \bv_{k',j'} \nonumber \\
& = & \sqrt{  \alpha_{k',j} \beta_{k',j} \Xc_{M-|\Uc_{k'}|+1}} + \sqrt{ \beta_{k',j} ( 1 - \alpha_{k',j})} \Gc,
\end{eqnarray}
where $\Xc_{m}$ denotes a chi-squared RV with $2m$ degrees of freedom and mean $m$, and $\Gc$ denotes a $\sim \Cc\Nc(0,1)$ RV. 
This follows from a well-known property of the pseudo-inverse of a Gaussian IID matrix with components $\sim \Cc\Nc(0,1)$ and dimension
$M \times |\Uc_{k'}|$, yielding the chi-squared term with $2(M-|\Uc_{k'}|+1)$ degrees of freedom. 

Let $k'$ denote a RRH for which pilot $q$ is trusted, and consider any precoding vector $\bv_{k',j'}$ 
where user $j'$ is associated to a different (trusted) pilot $q' \neq q$. 
By the orthogonality of $\bv_{k',j'}$ and $\widehat{\gv}_{k'}^{(q)}$  we have\footnote{For simplicity of notation, and since this is clear from the context, 
we use $\Gc$ and $\Xc$ to indicate Gaussian and chi-squared RVs, although these are different and mutually independent 
in different expressions.} 
\begin{eqnarray} \label{ZFBF-interf-residual}
\gv_{k',j}^\herm \bv_{k',j'} 
& = & \alpha_{k',j} (\widehat{\gv}_{k'}^{(q)})^\herm \bv_{k',j'} + \ev_{k',j}^\herm  \bv_{k',j'} =  \sqrt{ \beta_{k',j} ( 1 - \alpha_{k',j})} \Gc
\end{eqnarray}
Finally, for all RRH $k'$ for which pilot $q$ is not trusted, it follows that $\gv_{k',j}$ does not appear in any of the channel estimates 
$\{ \widehat{\gv}_{k'}^{(q')}\}$ used to compute the ZFBF precoding vectors.  Hence, the precoding vectors $\bv_{k',j'}$ of RRH $k'$ 
are simply unit vectors statistically independent of $\gv_{k',j}$, and we have
\begin{align} \label{conjbf-coef-noncoh}
\gv_{k',j}^\herm \bv_{k',j'} = \sqrt{\beta_{k',j}} \Gc .
\end{align}

Using the above results, we can determine each term appearing in the achievable spectral efficiency expression (\ref{rate1}). 
In particular, using  (\ref{ZFBF-coherent}), the expectation of the useful signal coefficient, i.e., of 
the first term in (\ref{DL}), can be written as
$\EE [ \mbox{useful} ] 
= \sum_{k \in \Ac_j}  \sqrt{\alpha_{k,j} \beta_{k,j}}  \EE[\sqrt{\Xc_{M - |\Uc_k|+1,k}}]$.
Since we are interested in the regime of $M \gg |\Uc_k|$ (massive MIMO regime), 
by the law of large numbers,  $\frac{1}{M - |\Uc_k|+1} \Xc_{M - |\Uc_k|+1} \rightarrow 1$ w.p.1. 
Hence, a simplified large-$M$ approximation yields
\begin{align} \label{useful-DL3}
\EE [ \mbox{useful} ] &= \sum_{k \in \Ac_j}  \sqrt{\alpha_{k,j} \beta_{k,j}}  \sqrt{M - |\Uc_k|+1} .
\end{align} 
Under this approximation, the variance of the useful signal term is given by 
\begin{equation} \label{useful-var}
{\rm Var}(\mbox{useful}) \approx \sum_{k \in \Ac_j} \beta_{k,j} ( 1 - \alpha_{k,j}) . 
\end{equation}
As far as multiuser interference is concerned, we distinguish 
between the non-copilot interference and the co-pilot interference. 
The non-copilot interference is due to all data streams transmitted by RRHs $k \in \overline{\Ac}_j^c$ as well as all data streams 
from RRHs $k \in \Ac_j \cup \widetilde{\Ac}_j^c$, except those set to users $j' \neq j$ also useing pilot $q$. 
We notice that the interference due to data streams sent by RRHs $k \in \Ac_j \cup \widetilde{\Ac}_j^c$ to UEs $j' \neq j$ 
generate terms of the type (\ref{ZFBF-interf-residual}). This yields the interference power 
\begin{equation} \label{residual-ZFBF}
\EE[|\mbox{residual ZF interference}|^2] = \sum_{k \in \Ac_j \cup \widetilde{\Ac}^c_j} \beta_{k,j} ( 1 - \alpha_{k,j}) (|\Uc_k| - 1) 
\end{equation}
The interference due to RRHs $k \in \overline{\Ac}_j^c$  yields terms of the type (\ref{conjbf-coef-noncoh}), thus 
\begin{equation} \label{non-copilot-ZFBF}
\EE[|\mbox{non-copilot interference}|^2] = \sum_{k \in \overline{\Ac}_j^c}  \beta_{k,j} |\Uc_k| 
\end{equation}
Finally, the copilot interference includes one term of the type of (\ref{ZFBF-coherent}) for each RRH $k \in \widetilde{\Ac}_j^c$. We have
\begin{eqnarray} \label{copilot-ZFBF}
\EE[|\mbox{copilot interference}|^2] 
& =  & \sum_{k \in \widetilde{\Ac}^c_j} \EE \left [ \left | \sqrt{\alpha_{k,j} \beta_{k,j} \Xc_{M-|\Uc_k|+1,k}} + \sqrt{ \beta_{k,j} ( 1 - \alpha_{k,j})} \Gc_k  \right |^2 \right ] \nonumber \\
& =  & \sum_{k \in \widetilde{\Ac}^c_j} \alpha_{k,j} \beta_{k,j} (M-|\Uc_k|+1) + \beta_{k,j} ( 1 - \alpha_{k,j})
\end{eqnarray}
Plugging (\ref{useful-DL3})-(\ref{copilot-ZFBF}) in (\ref{rate1}) yields the final expression for $C_j$ as in (\ref{rate-ZFBF}).


\section{Proof of (\ref{eq:PDF of Aun estimate})} \label{proof:PDF of Aun estimate}

Denote by $\nu$ the expected uncovered area fraction of general Borel set $B$ and 
${\1}(\cdot)$ the indicator function returning $1$ if its argument is true and $0$ otherwise. Then, we have
\begin{align}
 \Ebb \left[ \int_B \1 (  x \, \textnormal{is not covered}  )  \, \Dd x \right] &= 	  \int_B \Ebb \left[ \1 (  x \, \textnormal{is not covered}  )  \right]  \, \Dd x \label{eq:expect one} \\
 &= \int_B \Pbb \left(  x \, \textnormal{is not covered}  \right)  \, \Dd x \label{eq:expect two}
\end{align}
Invoking the void probability of a PPP in~(\ref{eq:expect two}) and dividing the resulting expression by $|B|$ gives the expected uncovered area fraction as
\begin{align} \label{eq:mean uncov area}
\nu = e^{ - \pi \lambdauq \Rout^2}
\end{align}
The main idea behind (\ref{eq:PDF of Aun estimate}) consists of approximating $f_\theta(\cdot)$ with two point masses at $\theta = 0$ and 
$\theta = 1$, and a uniform distribution for  $\theta \in [0,1]$.  Let us denote by $\pnot$ and $\pone$ the point masses of $\hat{f}_{\theta}(\cdot)$ 
at $\theta=0$ and $\theta=1$, respectively, and by $\pu$ the uniform distribution level 
of $\hat{f}_{\theta}(\cdot)$. Then, we have
\begin{align}
\pnot + \pu + \pone &= 1 \label{eq:pdf relation1} \\
\frac{\pu}{2} + \pone &= e^{ - \pi \lambdauq \Rout^2} \label{eq:pdf relation2} 
\end{align}
where (\ref{eq:pdf relation2}) follows by matching the mean with the value calculated in  (\ref{eq:mean uncov area}). 
Furthermore, the point mass at $\theta=1$ is equal to the probability that no other user is within distance $\Rin + \Rout$,  i.e.,
\begin{align} \label{eq:pone}
\pone &= e^{ - \pi \lambdauq (\Rin + \Rout)^2}.
\end{align}
Finally, (\ref{eq:pdf relation1}), (\ref{eq:pdf relation2}), and \eqref{eq:pone} determine the values of $p_0$, $p_1$ and $p_{\rm u}$, 
yielding (\ref{eq:PDF of Aun estimate}).

\section{Average Number of Users Per Cell} \label{proof: AvgNoUsersPercell}

Let us denote by $\pact$ the probability that a BS assigns pilot $q$ to one of its associated users. 
Then, the locations of BSs which also use pilot $q$ can be viewed as a thinned version of the original point process, 
denoted by $\Phibact^{(q)}$, whose density equals $\pact \lambdab$.
The thinning probability $\pact$, 
can computed as
\begin{align}
\pact 
&= \sum_{\ell=0}^{\infty} \Pbb \left( \, \textnormal{Pilot $q$ is assigned to a user in $\Uc_0$}  \, | |\Uc_0| = \ell  \right) \Pbb \left( |\Uc_0| = \ell \right)  \label{eq:pact step 1} \\
&= \sum_{\ell=0}^{\infty} \frac{\binom{L-1}{\ell-1}}{\binom{L}{\ell}}  \Pbb \left( |\Uc_0| = \ell \right)  \label{eq:pact step 11} \\
&= \sum_{\ell=0}^{\infty} \frac{\ell}{L}  \Pbb \left( |\Uc_0| = \ell \right) \label{eq:pact step 2} \\
&= \frac{1}{L} \Ebb \left[ |\Uc_0| \right]  \label{eq:pact step 3}
\end{align}

In turn, the expectation in (\ref{eq:pact step 3}) is written as
\begin{align}
\Ebb \left[ |\Uc_0| \right] 
&= \Ebb \left[ \min \left( |\Vcal_0|, \Np \right) \right] \label{eq: avg no of users 0} \\
&= \sum_{\ell=0}^{\Np  - 1} \ell \, \Pbb \left( |\Vcal_0| = \ell \right) +  \Np \, \Pbb \left( |\Vcal_0| \geq \Np \right) \label{eq: avg no of users 1}  \\
&= \sum_{\ell=0}^{\Np  - 1} \ell \, \Pbb \left( |\Vcal_0| = \ell \right) +  \Np \, \left( 1 - \sum_{\ell=0}^{\Np  - 1} \, \Pbb \left( |\Vcal_0| = \ell \right) \right) \label{eq: avg no of users 2}  \\
&= \Np + \sum_{\ell=0}^{\Np  - 1} (\ell - \Np) \, \Pbb \left( |\Vcal_0| = \ell \right) . \label{eq: avg no of users 3}
\end{align}
where the probability mass function of the number of points in $\Phiu \cap \Vcal_0$ where $\Vc_0$ is a Voronoi region of $\Phib$ 
is obtained by approximating the area of the Voronoi cell by a gamma-distributed random variable 
with a shape parameter $c = 3.575$ and a scale parameter $1/c \lambdab$\cite{ElSawy-stocgeo-tut-2017}, giving
\begin{align}
\Pbb \left( |\Vcal_0| = \ell \right) 
&= \frac{\Gamma(\ell+c)}{ \Gamma(\ell+1) \Gamma(c)} \frac{ (\lambdau)^{\ell} (c\lambdab )^c }{  (c \lambdab +  \lambdau)^{\ell+c} } \quad \ell=0, 1, 2, \ldots \label{eq:PMF no of poitns PPP}
\end{align}
Plugging (\ref{eq:PMF no of poitns PPP}) in (\ref{eq: avg no of users 3}) yields 
the sought value of $\Ebb \left[ |\Uc_0| \right]$,  which is further invoked in (\ref{eq:pact step 3}) to obtain (\ref{eq:pact cellular}).


\bibliographystyle{ieeebib}
\bibliography{jour_short,conf_short,refs_Ratheesh}

\end{document}